\documentclass[prb,aps,eqsecnum,twocolumn,amssymb,superscriptaddress,showpacs]{revtex4}

\usepackage{color}
\usepackage{graphicx}
\usepackage{longtable}
\usepackage{epsfig}
\usepackage{dcolumn}
\usepackage{amssymb}
\usepackage{epsfig}% Include figure files
\usepackage{graphicx}% Include figure files
\usepackage{dcolumn}% Align table columns on decimal point
\usepackage{bm}% bold math
\usepackage{curves}
\usepackage{epic}
\usepackage{wasysym}
\usepackage{subfigure}

\def\beq{\begin{equation}}
\def\eeq{\end{equation}}
\def\bea{\begin{eqnarray}}
\def\eea{\end{eqnarray}}

\def\nn{\nonumber}
\def\hst{\hspace{0.7cm}}

\begin{document}

\title{Magnetic properties of nanoscale compass-Heisenberg planar clusters}

\author {    Fabien Trousselet }
\affiliation{Max-Planck-Institut f\"ur Festk\"orperforschung,
             Heisenbergstrasse 1, D-70569 Stuttgart, Germany }

\author {    Andrzej M. Ole\'s }
\affiliation{Max-Planck-Institut f\"ur Festk\"orperforschung,
             Heisenbergstrasse 1, D-70569 Stuttgart, Germany }
\affiliation{Marian Smoluchowski Institute of Physics, Jagellonian
             University, Reymonta 4, PL-30059 Krak\'ow, Poland }

\author {    Peter Horsch }
\affiliation{Max-Planck-Institut f\"ur Festk\"orperforschung,
             Heisenbergstrasse 1, D-70569 Stuttgart, Germany }

\date{\today}

\begin{abstract}
We study a model of spins $1/2$ on a square lattice, generalizing the
quantum compass model via the addition of perturbing Heisenberg interactions
between nearest neighbors, and investigate its phase diagram and magnetic
excitations. This model has motivations both from the field of strongly
correlated systems with orbital degeneracy and from that of solid-state
based devices proposed for quantum computing.
We find that the high degeneracy of ground states of the compass model
is fragile and changes into twofold degenerate ground states
for any finite amplitude of Heisenberg coupling.
By computing the spin structure factors of finite clusters with
L\'anczos diagonalization, we evidence a rich variety of phases
characterized by $\mathbb{Z}_2$ symmetry, that are either ferromagnetic,
$C$-type antiferromagnetic, or of N\'eel type,
and analyze the effects of quantum fluctuations on phase boundaries.
In the ordered phases the anisotropy of compass interactions leads to
a finite excitation gap to spin waves.
We show that for small nanoscale clusters with large anisotropy gap the
lowest excitations are column-flip excitations that emerge due to
Heisenberg perturbing interactions from the manifold of
degenerate ground states of the
compass model. We derive an effective one-dimensional XYZ model which
faithfully reproduces the exact structure of these excited states
and elucidates their microscopic origin.
The low energy column-flip or compass-type excitations are robust
against decoherence processes and are therefore well designed for
storing information in quantum computing. We also point out that the
dipolar interactions between nitrogen-vacancy centers forming a
rectangular lattice in a diamond matrix may permit a solid-state
realization of the anisotropic compass-Heisenberg model.

\end{abstract}

\pacs{75.10.Jm, 03.65.Ud, 05.30.Rt, 64.70.Tg}

\maketitle

\section{Introduction}
\label{sec:int}

\begin{figure}[t!]
\begin{center}
\includegraphics[width=3.4cm]{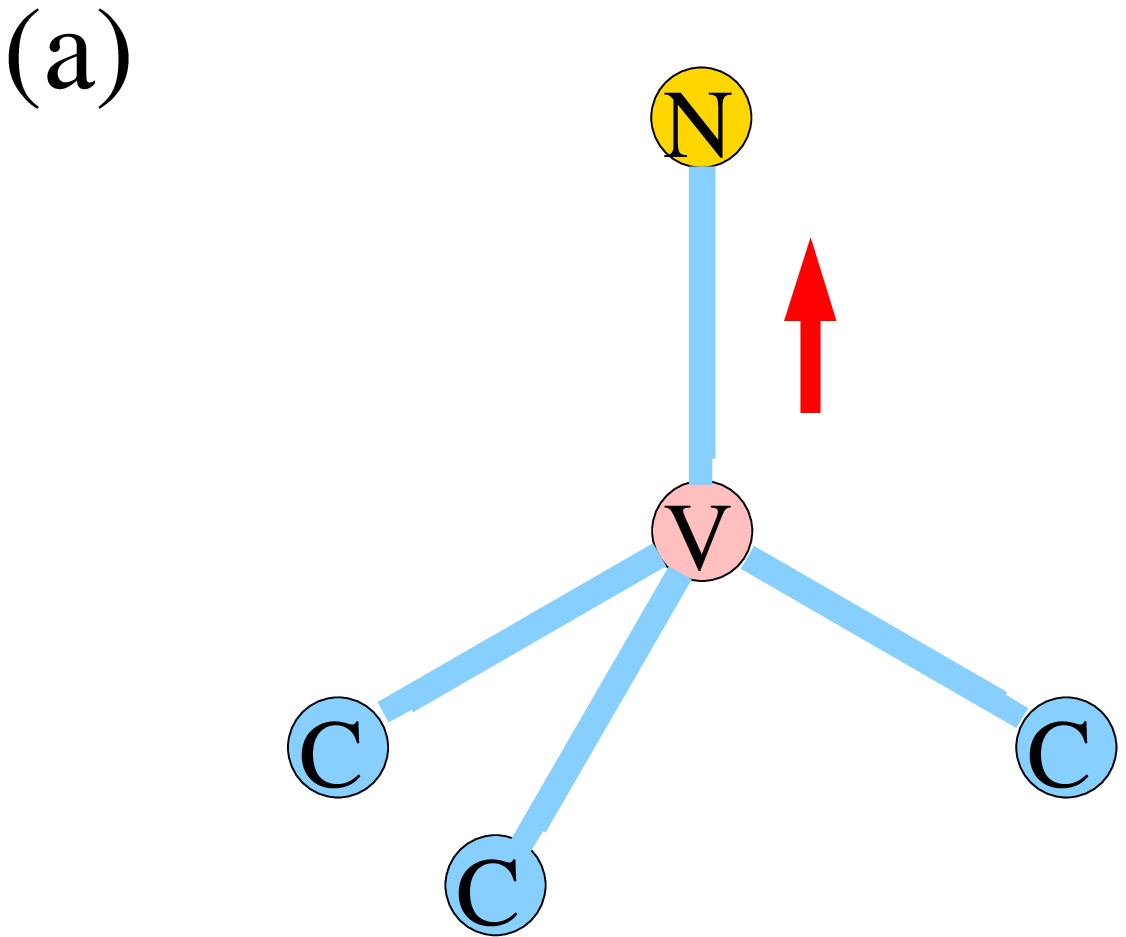}
\hspace{0.8cm}
\includegraphics[width=3.6cm]{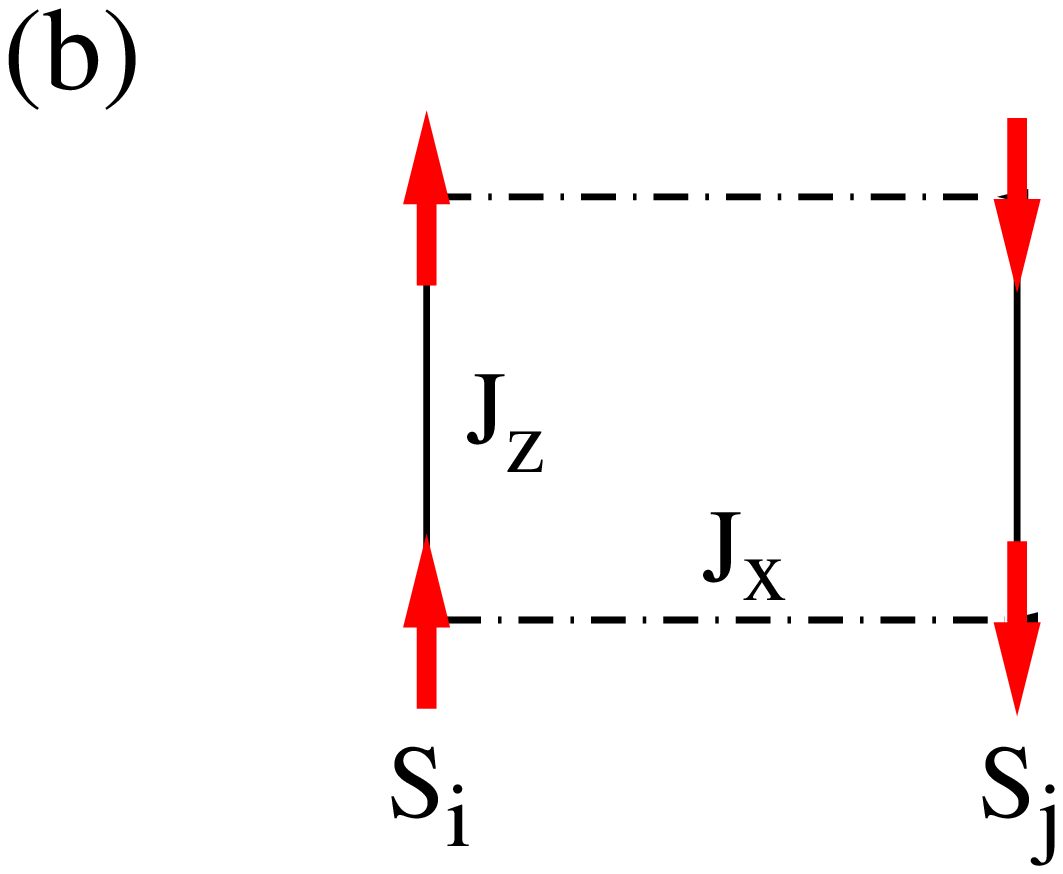}\\
\includegraphics[width=3.3cm]{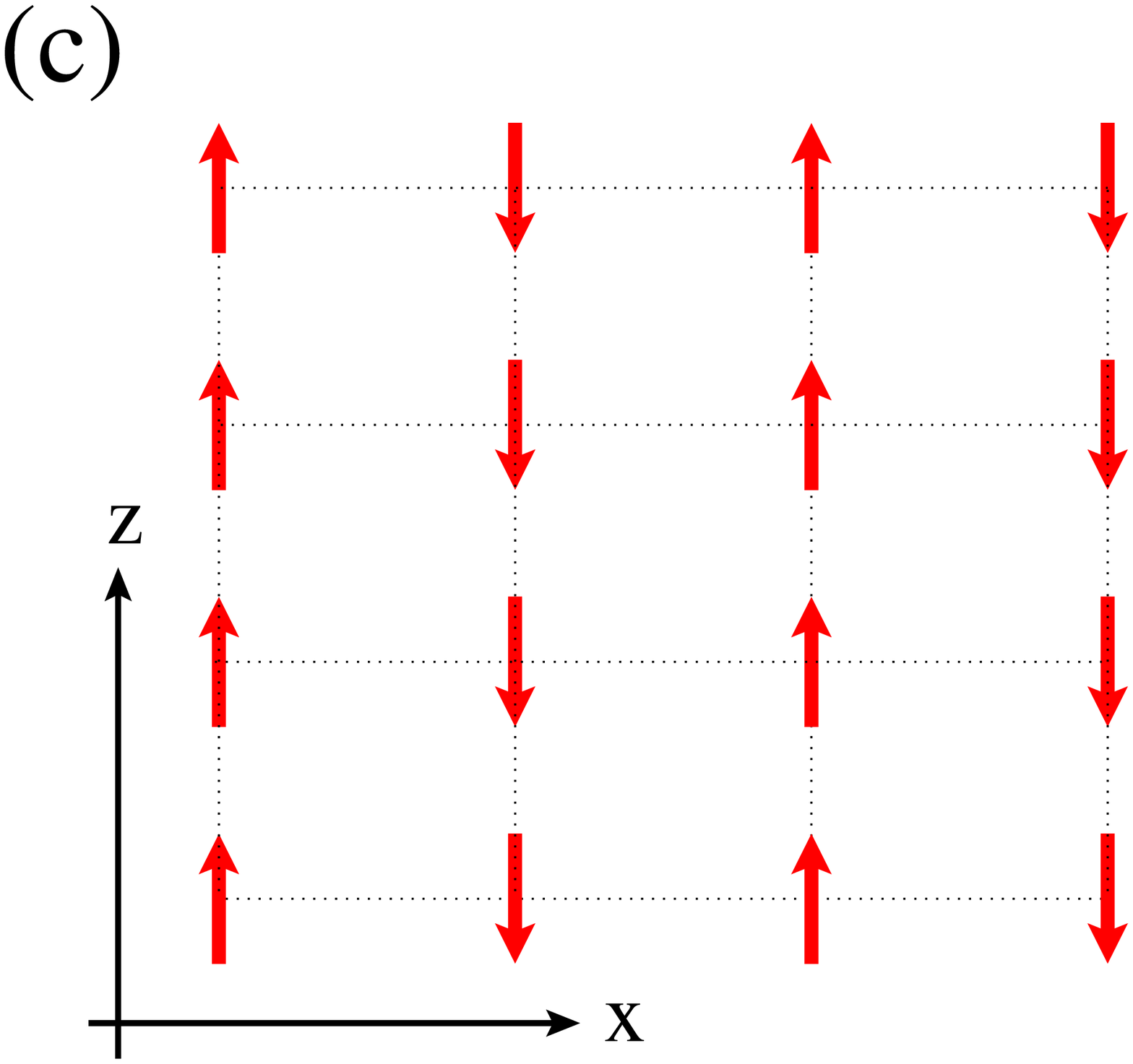}
\hspace{1.0cm}
\includegraphics[width=3.7cm]{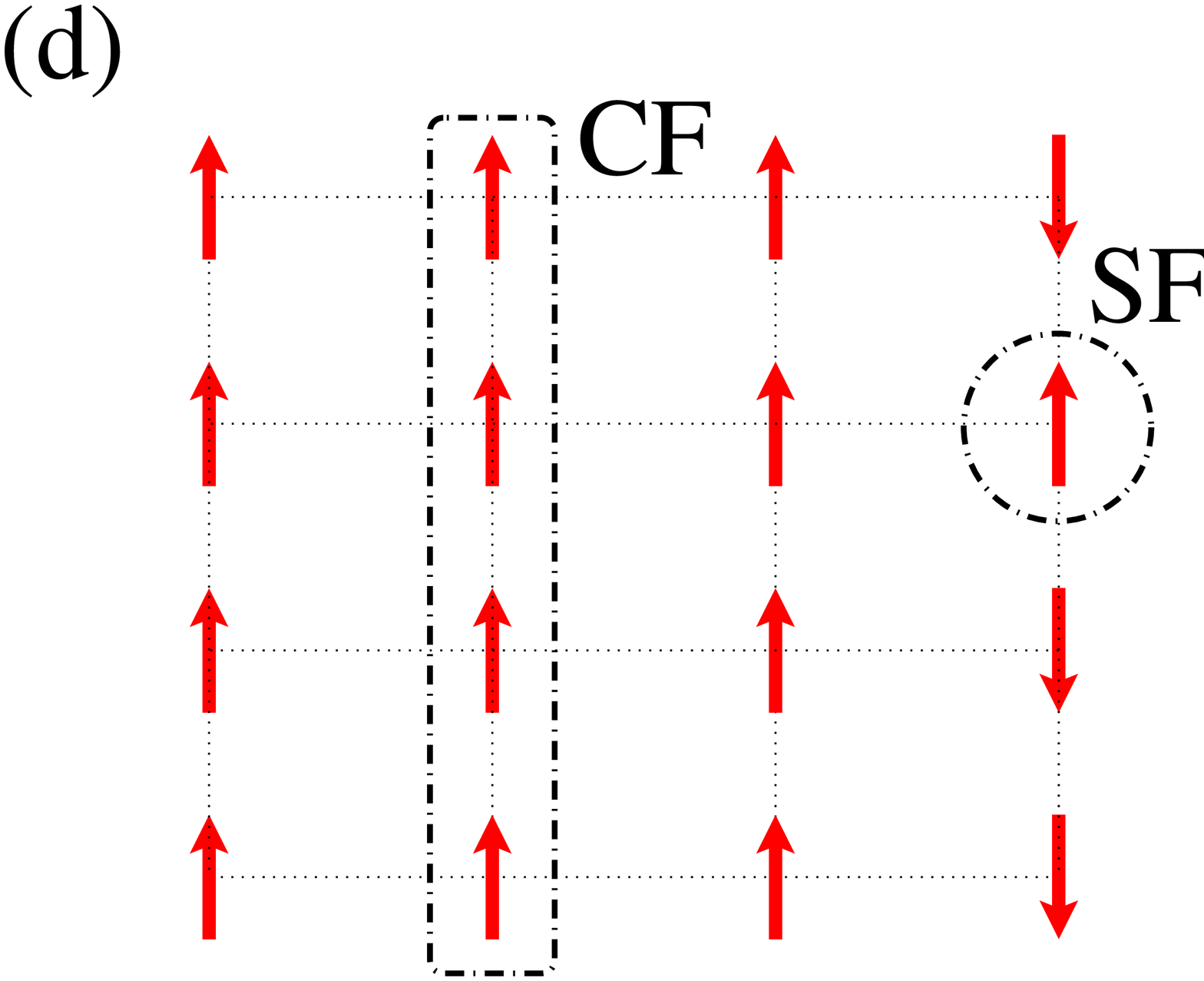}\\
\caption{\label{NVc}(Color online)
(a) Structure of an NV center in a diamond matrix,
which can be described by an effective spin;
(b) four NV centers forming
a rectangle are controlled by dipolar interactions between the associated
spins $\{\vec{S}_i\}$.  Dipole interaction $H_{\rm dip}$  involves
compass-type terms of various amplitudes $\{J_x,J_z\}$, depending on the
pair considered, and Heisenberg interactions.
(c) Ground state of anisotropic compass-Heisenberg model represented by
the spins of NV centers and arranged in a rectangular array.
(d) Distinct low-energy excitations of this system: a column flip (CF)
reverses the spins of a whole column, while a spin flip (SF) reverses a
single spin, contributing to spin-wave excitations.}
\end{center}
\end{figure}

Frustrated quantum magnetism belongs to the very active research areas
in condensed matter theory. Frustration is one of the simplest concepts
in physics with far reaching consequences.\cite{Nor09,Hon11} It is well known
that antiferromagnetic (AF) exchange for three spins on a triangle is
geometrically frustrated, both in classical Ising and in quantum
Heisenberg models. On a two-dimensional (2D) square lattice, frustration
typically involves interactions between further neighbors competing with
those between nearest neighbors, but it can also occur with only the
latter ones: for instance when, compared to the Ising model, the sign of
spin exchange along every second column is reversed. The resulting model,
called \textit{fully frustrated Ising model}, \cite{Vil77} is
exactly solvable, with a phase transition at a lower temperature than
that of the 2D Ising model \cite{Lon80} --- the low temperature phase,
with extensive entropy due to frustration, is described
in terms of dimer coverings of the dual lattice.\cite{Moe01} A quantum
analog of this classical frustrated model on a square lattice is the 2D
quantum compass model (QCM),\cite{Kho03}
where interactions couple either $S_i^x$ or $S_i^z$ components of nearest
neighbor $S=1/2$ spins, depending on the spatial bond direction $x$ or $z$
respectively. When the associated exchange constants $J_x$ and $J_z$
[see Fig.~\ref{NVc}(b)]
have different values, these two spin components are nonequivalent. Yet
even otherwise, and in spite of its $(2+1)$ dimensionality, this model
displays a finite-temperature phase transition of the 2D Ising universality
class, \cite{Wen08} but the symmetry broken phase at low temperature is
characterized by high ground state degeneracy in the thermodynamic limit.
\cite{Dor05}

One can interpolate between the 2D QCM and Ising models by modifying
continuously the spin components coupled on the bonds along two
distinct lattice directions.\cite{Cin10} This allows one to highlight
that the QCM is closely related to orbital physics, where the exchange
interactions are directional.\cite{vdB04} In fact, one finds a 2D
superexchange model for $e_g$ orbitals as an intermediate model when
the interactions are
gradually modified from the classical Ising model toward the QCM.
While the frustration of interactions is clearly weaker in the $e_g$
orbital model than in the QCM, the latter may be considered
as a generic description of frustrated directional orbital interactions
which arise for strongly correlated electrons in transition metal
oxides with partly filled degenerate $3d$ orbitals,
and is realized for instance in manganites.\cite{Lia11}
In these systems the orbital degrees of freedom play a crucial role in
determining ground states with coexisting magnetic and orbital order,
described by spin-orbital superexchange.
\cite{Kug82,Fei97,Nag00,Ole05,Kha05,Hfm,Ole12}

The orbital interactions are intrinsically frustrated \cite{Fei97}
as they have low symmetry in pseudospin space representing the orbital
degrees of freedom.
Typically the symmetry is that of the lattice due to the shape
of $3d$ wave functions, and not the SU(2) symmetry typical for spin
exchange interactions. Although frustration is at its maximum
in three-dimensional (3D) models and it was concluded from the
high-temperature expansion that a phase
transition to the symmetry-broken states does not occur,\cite{Oit11}
recent Monte-Carlo simulations have
evidenced symmetry-broken phases at low temperatures
both in $e_g$ and $t_{2g}$ orbital models.\cite{vRy10,Wen11}
This contrasts with a formal analog of the QCM defined on the honeycomb
lattice, the Kitaev model,\cite{Kit06} for which an exact solution
evidenced a spin liquid ground state. In fact, both models can describe
orbitally degenerate Mott insulators in the limit of strong spin-orbit
coupling \cite{Jac09} which selects a low-energy doublet at each
transition metal ion represented by a pseudospin-$1/2$ variable ---
which model is actually relevant depends on the geometry of the system.
However, realistic orbital models are more
involved,\cite{vdB04} {\it inter alia\/} due to non-conservation of the
orbital quantum numbers that follows both from hybridization processes
with oxygen orbitals in an oxide and from the structure of charge
excitations
controlled by Hund's exchange in orbital degenerate systems. The QCM was
designed to avoid all these complications and to address a paradigm of
intrinsic frustration due to directional conflicting interactions.

Another motivation for introducing and investigating the 2D QCM comes
from the field of quantum computing.\cite{Nie10,Ben06} Recent progress
includes proposals for the optimal choice of protected qubits.\cite{Jon12}
Several realizations of computing devices with protected qubits have been
proposed in various contexts:
(i) in Josephson junction arrays,\cite{Dou05,Gla09} as well as
(ii) with polar molecules, or
(iii) with systems of trapped ions in optical lattices.\cite{Mil07}
In all these cases the QCM provides the generic description of
interacting spins.

In general, in order to construct a device which could serve for
information storage, a manifold of degenerate states is required,
\cite{Compu} and these degeneracies
should be stable against noise and other small perturbations\cite{Dou05}
thanks to particular symmetries of the Hamiltonian. This is actually the
case in the quantum compass model, where two types of symmetry operations
described by operators $P_i$ and $Q_j$ commute with the Hamiltonian
but not with each other, see Sec. II. As a
result the eigenstates of the system are characterized by related
integrals of motion, and concerning the ground state, by a hidden dimer-dimer
symmetry.\cite{Brz10} More importantly, an exact twofold
degeneracy of all quantum levels was evidenced on finite clusters of
arbitrary size, being of advantage for quantum information.\cite{Dou05}
These degeneracies are, thanks to the non-local nature of
operators $P_i$ and $Q_j$, robust to local perturbations; in consequence qubits
defined by a realization of the QCM are expected to be protected against noise,
so that this model is of prime interest for quantum computing.

The QCM has thus an interdisciplinary character as it plays an important role
in the modeling not only of correlated transition metal oxides, but also
of protected qubits for quantum computations.
An intriguing question important in all these contexts and asked shortly
after the QCM was introduced concerns the nature of a quantum
phase transition (QPT) that occurs when anisotropic interactions are
varied through the isotropic point, also called the compass point. A
first order transition between two
distinct phases with directional ordering, along either rows or columns,
was suggested by L\'anczos diagonalization and Green's function
Monte-Carlo simulations for
finite clusters,\cite{Dor05} and later confirmed using a projected
entangled-pair state algorithm.\cite{Oru09}
At this transition, a discrete symmetry in spin space is spontaneously broken
since the frustrated interactions along two different directions are
equivalent and the spin orientation follows one of them. In terms of broken
symmetries this transition is remarkably similar to the first
order QPT found at $J_x=J_z$
in the exact solution of the one-dimensional (1D) QCM,
\cite{Brz07} or a compass ladder,\cite{Brz09} where two different types
of order stem from the invariant subspaces of the 1D model. This
suggests that a similar mechanism may operate also in two dimensions.

Particularly in the context of proposed realizations of quantum computing
devices based on finite clusters with compass-like spin interactions,
a fundamental question to ask is how the highly degenerate ground states
\cite{Dor05} are modified when a small perturbation occurs.
We argue that Heisenberg interactions between nearest neighbor spins
stand for a class of perturbations to the compass terms which are typical
in solid state systems --- for instance a Hamiltonian with compass and
Heisenberg terms would describe exchange processes in some
Mott insulators with strong spin-orbit coupling and 180-degree bonds.
\cite{Jac09} In a broader perspective we study in this work
the effect of Heisenberg perturbations, by considering a
generalization of the QCM called the
\textit{compass-Heisenberg} (CH) model.\cite{Tro10}

We find that the high degeneracy of ground states in the thermodynamic
limit (TL) is removed by Heisenberg terms of arbitrarily small amplitude,
and various magnetically ordered phases arise, with a preferred spin
direction related to the ordered pattern. In macroscopic systems the
lowest energy excitations are thus gapped spin waves;
on nanoclusters however, for small enough Heisenberg amplitude, another
type of excitations can be of lower energy than spin waves: these are the
\textit{column-flip excitations}, from the ordered ground states selected
by small Heisenberg terms to the many other eigenstates of the low-energy
manifold minimizing the energy of dominating compass interactions.
The column-flip excitations
are robust with respect to decay into spin waves, and preserve an original
multiplet structure which can be captured by an adapted effective model;
this analysis leads us to propose that these excitations could be used in
a novel type of solid-state-based quantum computing scheme
in a regime of moderate Heisenberg interactions.

In particular we find in the frame of the CH model
that the compass point ($J_z=J_x$, $I=0$) appears as a quadri-critical
point where four distinct phases with $\mathbb{Z}_2$ symmetry
meet in the plane spanned by two parameters, $J_x/J_z$ and $I/J_z$, 
characterizing the compass- and the Heisenberg couplings. We note that the 
transitions between arbitrary two phases $A$ and $A'$ related by a duality 
transformation are continuous
transitions for finite system size, while they appear as first order
transitions in the TL.\cite{Oru09} Here we find that also the
transitions between phases $A$ and $B$ belonging to
distinct $\mathbb{Z}_2$ symmetries show a similar behavior.
Remarkably, these transitions are characterized by the softening of certain
columnar excitations rather than of spin waves.

Recent experimental developments on arrays of nitrogen-vacancy (NV)
centers, constituting point-like defects in a diamond matrix,
\cite{Gae06,Neu10} may bring a further motivation to the study of a
model with coexisting compass and Heisenberg interactions. Indeed these
defects can be effectively described by quantum spins
$\vec{S}_i$,\cite{1vs12} coupled (under certain
conditions) predominantly by the dipolar interactions \cite{vvl} of the form:
\beq
H_{\rm dip} = \sum_{\langle ij\rangle\parallel\gamma}
\frac{C}{r_{ij}^3} \left(\vec{S}_i\cdot\vec{S}_j
-3S_i^\gamma S_j^\gamma\right),
\label{eq:Hdip}
\eeq
where the $\gamma$ axis in spin space is along the spatial direction
connecting spins $i$ and $j$. These interactions are
long-ranged, but rapidly decaying with distance; if defects sit on sites of
a rectangular cluster, the (dominant) interactions between nearest neighbors
are a sum of Heisenberg-like and compass-like terms.\cite{Tro10}
Beyond the nature of couplings, an aspect which must be taken into
consideration in this context is a possible splitting of
energy levels for a single NV center. One can {\it a priori\/} consider a
situation where these splittings are small before the typical energy scale of
dipolar couplings; alternatively, other possible realizations of
dipolar-coupled spin arrays are conceivable (with e.g. $I_n=1/2$ nuclear
spins, in a layered crystal with an orthorhombic
unit cell and negligible effects of hyperfine coupling to
electron spins). We will show below that
in such systems the lowest energy excitations can consist
of reversing entire columns of spins,
and these
could be used for encoding protected qubits, see Fig.~\ref{NVc}.

The paper is organized as follows. In Sec.~\ref{sec:chm} we introduce the
CH model and state the problem of frustrated interactions and possible
QPTs. There are two variants: the {\it ferromagnetic} (FM) and the
{\it antiferromagnetic} (AF) CH model. We first focus on the AF CH model,
and present selected data for the spin structure factors in
Sec.~\ref{sec:s(q)} and show that long-range order is
induced by arbitrarily small Heisenberg interactions.
The full phase diagram of the CH model is
presented in Sec.~\ref{sec:phd} --- there, we provide
evidence that some phase transitions occur for the same interaction
parameters as in the classical CH model, while other transition
lines are affected by quantum fluctuations, see Secs.~\ref{sec:sym}
and \ref{sec:pt}. Next we analyze the features of the FM CH model and
discuss its phase diagram in Sec. \ref{sec:fm}. Spin wave excitations
are derived and discussed for different phases in Sec.~\ref{sec:sw}.
We turn then to the analysis of the lowest energy states of finite
clusters and show in Sec.~\ref{sec:cex} that:
(i) the ground state and the low energy excitations are very well
described by an effective 1D model which captures the essential
parameter dependence of columnar (i.e., column-flip) excitations
characteristic of the compass-Heisenberg model; and
(ii) there exists
a parameter range where the column-flip excitations are the lowest energy
excitations and cannot decay into spin waves. The paper is summarized in
Sec.~\ref{sec:summa}, where open issues and possible extensions of this
work are also discussed.

\section{Compass-Heisenberg model}
\label{sec:chm}

We consider a model of spins $S=1/2$ on the square lattice, with axes in
the $ab$ plane, labeled here $x$ and $z$ after the interacting spin
components in the QCM. The nearest neighbor interactions are of two types:
(i) frustrated compass interactions of amplitudes $J_x$ and $J_z$, and
(ii) Heisenberg interactions with an exchange $I$.
While the Heisenberg interaction is isotropic in spin space
and bond-independent,
the compass interactions depend on the bond direction.
On bonds along the $x$-axis the $x$-components of spins are coupled by
terms $\sigma^x_{i,j}\sigma^x_{i,j+1}$
(we label the sites in a 2D cluster by two indices $\{i,j\}$),
and on bonds along the $z$-axis
the coupling concerns the $z$-components, being of the form
$\sigma^z_{i,j}\sigma^z_{i+1,j}$. For convenience we use here Pauli
matrices ${\vec\sigma}_{\vec r}\equiv\{\sigma^x_{\vec r}, \sigma^y_{\vec
  r},\sigma^z_{\vec r}\}$ with $\vec r=(i,j)$ such that
$\sigma^z_{\vec r}=\pm 1$ in the $\sigma^z$ basis.

The CH Hamiltonian reads: \cite{Tro10}
\begin{eqnarray}
{\cal H}&=& \sum_{i,j} \left( J_x\sigma^x_{i,j} \sigma^x_{i,j+1}
        + J_z\sigma^z_{i,j} \sigma^z_{i+1,j}\right) \nonumber \\
 &+& I\sum_{i,j}\left( \vec{\sigma}_{i,j}\cdot\vec{\sigma}_{i,j+1}
             +\vec{\sigma}_{i,j}\cdot\vec{\sigma}_{i+1,j}\right)\,.
\label{ham}
\end{eqnarray}
Here the sums over $i$ and $j$ run over the intervals
$\mathbb{[}1,L_z\mathbb{]}$ and $\mathbb{[}1,L_x\mathbb{]}$ consistent
with either periodic boundary conditions (PBC) or open boundary
conditions (OBC). We consider rectangular clusters with
$N=L_x\times L_z$ spins and $L_x,L_z \le 6$ for both PBC and OBC.
Another type of clusters (considered only with PBC) are clusters tilted
by $\pi/4$ w.r.t. the previous ones and containing $N=2L^2$ spins
(for $L=3,4$).\cite{Leu95}

\begin{figure}[t!]
\includegraphics[width=6cm]{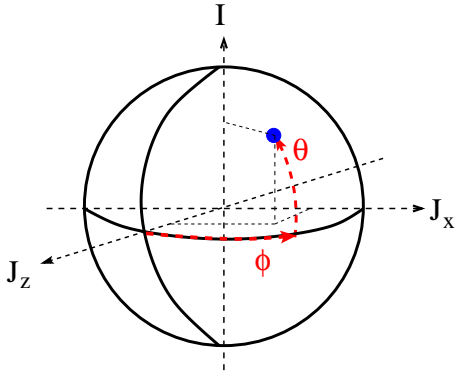}\\
\caption{(Color online) Parametrization of the compass-Heisenberg model
Eq.~(\ref{ham}) by two angles $\{\phi,\theta\}$.
A point in parameter space is indicated by the (blue) dot on the sphere;
interaction parameters $I$, $J_x$, and $J_z$ correspond to its cartesian
coordinates on the respective axes, while angles $\theta$ and
$\phi$ are its spherical coordinates.}
\label{fig:sphere}
\end{figure}

\begin{figure}[t!]
\includegraphics[width=6cm]{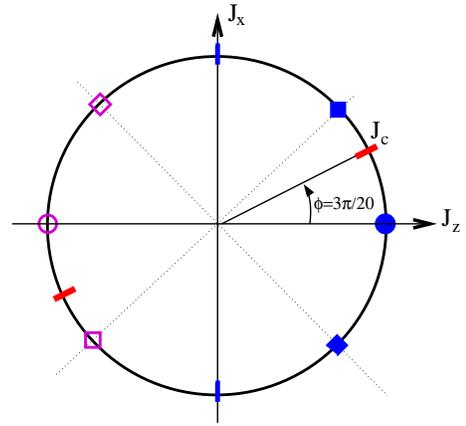}\\
\caption{\label{circomp}(Color online)
Two equivalent parametrizations of the compass model by either
couplings $J_z$ and $J_x$ or by
the radius $J_c$ and angle $\phi$
defined by Eqs. (\ref{Jc}) and (\ref{phi}), respectively.
On this circle, isotropic compass points $|J_x|=|J_z|$ are
shown by filled (blue) square and diamond for the
model with AF $J_z>0$ interactions and by empty
(purple) square and diamond for the model with FM
$J_z<0$, while Ising points are indicated by
dots ($J_x=0$) or vertical bars ($J_z=0$).
The points $\phi=3\pi/20$ and $\phi=23\pi/20$, frequently
considered hereafter, are also indicated by radial (red) bars.}
\label{fig:circle}
\end{figure}

The structure of eigenstates of $\cal{H}$ depends only on the relative
amplitude of parameters $J_z$, $J_x$ and $I$ in Eq.~(\ref{ham}). Thus, the
total space of interaction parameters may be characterized by a point on the
spherical surface parametrized by angles $\{\phi,\theta\}$,
see Fig.~\ref{fig:sphere}. The compass interactions are then
described by the related global interaction strength,
\begin{equation}
J_c=\sqrt{J_z^2+J_x^2},
\label{Jc}
\end{equation}
and the angle $\phi$ determines the exchange constants,
\begin{equation}
\label{phi}
J_z=J_c\cos\phi \,, \hspace{1cm} J_x=J_c\sin\phi \,,
\end{equation}
that are represented by a point in the $(J_x,J_z)$ plane, see
Fig.~\ref{fig:circle}. Using this parametrization the Heisenberg
interaction $I$ is given by the angle $\theta$:
\begin{equation}
\tan\theta=\frac{I}{J_c}\,.
\end{equation}
In the following we shall denote by \textit{antiferromagnetic
  compass-Heisenberg} (AF CH) model the case where $J_z>0$, and by
\textit{ferromagnetic compass-Heisenberg} (FM CH) model the case $J_z<0$.

We introduce certain non-local operators playing a central role in the QCM
(and, as we will see, in the CH model), and defined either on rows $i$
or on columns $j$ of the considered clusters, by:
\bea
P_i= \prod_j \sigma^z_{i,j}, \\
Q_j= \prod_i \sigma^x_{i,j}.
\label{piqj}
\eea
Here $Q_j$ rotates all spins in the column $j$
from up- to down-orientation or vice versa --- this operation is called column
flip (CF), see Fig~\ref{NVc}(d). Similarly, $P_i$ rotates a whole row of spins
pointing along $\pm x$ direction into $\mp x$ direction in spin-space. In the
QCM ($I=0$) these operators are known \cite{Dor05,Dou05} to commute
with the compass Hamiltonian ${\cal{H}}_{I=0}$
(i.e., $[P_i,{\cal{H}}_{I=0}]=0$, $[Q_j,{\cal{H}}_{I=0}]=0$) and ---
restricting now to the first type of (untilted) clusters --- they
anticommute ($\{P_i,Q_j\}=0$), accounting for an exact twofold degeneracy.

In presence of Heisenberg interactions, the commutators
above become non-zero. To evaluate those, it is useful to consider
separately two terms $\propto I$, complementary to each other in
Eq.~(\ref{ham}):
the first term ${\cal{H}}_I^r$ acts on horizontal bonds (rows),
while the second term ${\cal{H}}_I^c$ acts on vertical bonds (columns).
It is now straightforward to show that $Q_j$ commutes with the columnar
interaction, i.e., $[Q_j,{\cal{H}}_I^c]=0$, but not with the Heisenberg
interactions on the rows
\begin{equation}
[Q_j,{\cal{H}}_I^r]  =  2 I Q_j  \sum_{i,\alpha\in \{y,z\}}\!
\left(\sigma^{\alpha}_{i,j-1}\sigma^{\alpha}_{i,j}
+\sigma^{\alpha}_{i,j}\sigma^{\alpha}_{i,j+1}\right).
\label{comut}
\end{equation}
Hence the column $j$ is here coupled to the left ($j-1$) and the right ($j+1$)
column by $\sigma^z \sigma^z$ and $\sigma^y \sigma^y$ components of 
Heisenberg interactions.

These operators are related to a formalism which allows one to
understand the high ground state degeneracy of the QCM in the TL,
and which we will briefly describe here. We
consider a situation with strong anisotropy $|J_z|\gg |J_x|$, which calls for
a perturbative treatment of $J_x$ couplings. The unperturbed
Hamiltonian contains only the $J_z$ compass couplings, which select, on a
$(L_x,L_z)$ rectangular cluster, $2^{L_x}$ \textit{columnar states}.
These states, where in each column all spins point along the
same axis $z$ --- aligned either ferromagnetically or antiferromagnetically
depending on the sign of $J_z$ ---- can be labeled using pseudospin variables
$\vec{\tau}_j$, with $1\le j\le L_x$.
For a given columnar state, a given column $j$ will be described by an
eigenstate of the pseudospin operator $\tau_j^z$, either
$\tau_j^z|\uparrow_j\rangle=+|\uparrow_j\rangle$ or
$\tau_j^z|\downarrow_j\rangle=-|\downarrow_j\rangle$,
depending on whether the spin at a reference site $(1,j)$ has the orientation
up or down respectively; orientations of other spins in the column follow
from its FM ($J_z<0$) or AF ($J_z>0$) long-range ordered nature.
In both cases, the operators $\tau_j^x$ and $\tau_j^y$ flip all spins
of the column $j$ with amplitudes given by the respective Pauli
matrices. The operator $\tau_j^x$ has actually the same action on a
column $j$ as the operator $Q_j = \prod_{i} \sigma^x_{i,j}$, with the
only difference that $\tau_j^x$ is defined only in the subspace
generated by columnar states. The action of $\tau_j^x$ operators on a 
reference columnar state defines \textit{column-flip} excitations,
which will be analyzed in  Sec.~\ref{sec:cex} and correspond qualitatively to
flipping all spins in a column of a finite cluster [see Fig.~\ref{NVc}(d)].
They are well defined when perturbing interactions favor a particular
columnar pattern in the ground state, and we will see that
this is typical for the CH model.

Within the QCM (for $I=0$), the perturbation theory describing the 
effects of small couplings $\propto J_x$ acts in the subspace 
of \textit{columnar states}.\cite{Dou05} We recall the expression of the 
effective Hamiltonian obtained at leading order: \cite{Dou05,Dor05}
\bea
H^{(0)}_{\rm col}= -J_{\rm col} \sum_{j=1 \ldots L_x} \tau_j^x \tau_{j+1}^x .
\eea
Here the effective coupling constant $J_{\rm col}$ describing the flip of a
whole column is obtained at order $L_z$ in perturbation theory:
\bea
J_{\rm col}=2 L_z 2^{L_z} \gamma^{(')}_{L_z} J_z
\left|\frac{J_x}{8J_z}\right|^{L_z}.
\label{eq:efJx}
\eea
The coefficient $\gamma^{(')}_{L_z}$ depends on boundary conditions
(with or without a prime for OBC or PBC, respectively) and on the column
length; it can be determined by considering all processes flipping two
neighboring columns $i$ and $(i+1)$, by $L_z$ successive actions of
perturbing terms $J_x\sigma_{i,j}^x\sigma_{i,j+1}^x$.

Assuming PBC, the
excitation energies of intermediate states during such $L_z$-th order
processes are integer multiples of the quantity $8|J_z|$ which appears
in Eq.~(\ref{eq:efJx}). The counting of these processes, weighted by a
factor depending on the excitation energies at each step of each
process, is a combinatorial problem which, to our knowledge, does not
have a general analytic solution; however for small $L$ the exact values
of $\gamma^{(')}_L$, or equivalently of $c^{(')}_L=L 2^{L-2} \gamma^{(')}_L$,
are easily computable. As examples we give here $\gamma_4=5/4$,
$\gamma_5=7/4$ and $\gamma_6=29/12$. In the case of
OBC, the number of processes flipping two neighboring columns of $L$
sites is the same as for PBC, but the excitation energies at some
intermediate steps may be lower than in the periodic case so that
$c'_L \geq c_L$, e.g. for $L=4$ one has $c'_4=8c_4/5$.
One can even remark that with PBC
$\gamma_L \geq 1$, by noticing that
there are exactly $L\times 2^{L-2}$ column-flipping processes for which
the excited energy at each step is minimal, i.e., $8|J_z|$ (these are
the processes where two successive actions of perturbing terms occur on
bonds distant by 1 unit along the $z$ axis).

A scaling law for the size-dependence of $c_L$, or equivalently
of $J_{\rm col}$, was given in Ref.~\onlinecite{Dor05}, indicating that the
latter vanishes exponentially with increasing $L_z$ --- in the compass model
with $|J_x|<|J_z|$ this yields precisely the $2^{L_x}$-fold ground state
degeneracy in the TL. The isotropic case $J_x=J_z$ has a higher ground state
degeneracy $2^{L_x}+2^{L_z}$ in the TL, which can be deduced from
similar arguments.

As mentioned before, the compass model itself is characterized by a high
level of frustration between $\sigma^z \sigma^z$ and $\sigma^x \sigma^x$
interactions, independent of the sign of the associated amplitudes.
From that perspective, the introduction of perturbing Heisenberg
interactions seems to increase the degree of frustration in the model,
e.g. in a case where they are of sign opposite to that of dominant
compass interactions. The ordered patterns favored in this case, if ever,
are expected to differ from those selected for dominant Heisenberg
interactions, i.e., $|I| \gg J_c$ - in the former case, a ground state
minimizing energy both of dominant compass interactions (on either
vertical or horizontal bonds, depending on the sign of $|J_z|-|J_x|$)
and of Heisenberg interactions (on other bonds) can be selected; while
in the latter, a FM or N\'eel order is expected with an easy axis
selected by compass couplings. Thus, besides the question of whether the
exotic, semi-disordered ground states characteristic of the compass
model can actually exist in presence of Heisenberg couplings with small
amplitudes, one can focus in this model on the determination of the
phase diagram, with multiple phase transitions between the more
conventional FM or N\'eel phases, and more exotic $C$-type AF
phases, with FM order along one axis and AF along the other. The
characterization of these phases is the subject of the next chapter.

\section{Spin structure factors}
\label{sec:s(q)}

In this Section we address the following central question: What happens
to the macroscopic $2^L$-fold ground-state degeneracy of the anisotropic
QCM in the presence of Heisenberg perturbations? We show that in the
most general case, where compass coupling strengths $|J_x|$ and $|J_z|$
are not equal and where the Heisenberg coupling strength $|I|$ is
finite, the ground state $|\Psi_0\rangle$ is characterized by
long-range spin order with a certain easy axis.
This is evidenced by spin structure factors $S^\alpha(\vec q)$,
defined for each orthogonal spin component $\alpha \in \{x,y,z\}$ as:
\beq
S^\alpha(\vec q) = \frac{1}{N} \sum_{\vec r,\vec s} e^{i\vec{q}\cdot (\vec{r}
- \vec{s})} \langle \Psi_0| \sigma^\alpha_{\vec r} \sigma^\alpha_{\vec s}
|\Psi_0 \rangle.
\label{saq}
\eeq
For given interaction parameters, a peak in $S^\alpha(\vec q)$ at a single
momentum $\vec q_0$ and component $\alpha$ signals an ordering of spins with
a finite component along the $\alpha$ axis and a modulation period given by
$\vec q_0$.

\begin{figure}[t!]
\begin{center}
\includegraphics[width=7.8cm]{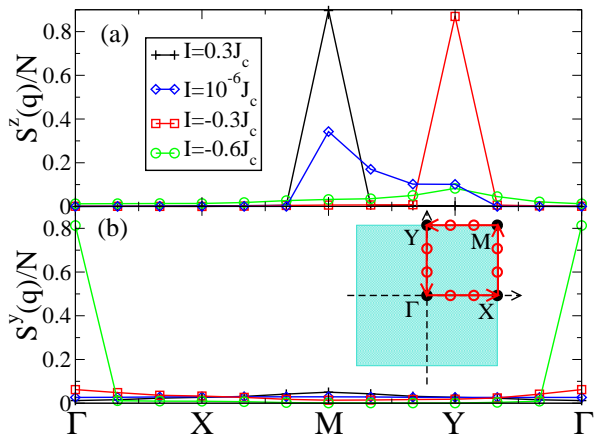}\\
\end{center}
\caption{(Color online) Structure factors (a) $S^z(q)$ and (b) $S^y(q)$
(divided by $N$) obtained for the AF CH model with the
$N=36$ cluster (symbols). The lines are guides to the eye.
Parameters: $\phi=\pi/10$; the data correspond to four different values
of $I/J_c$: $0.3$ ($G_z$ phase), $+10^{-6}$,
$-0.3$ ($C'_z$ phase) and $-0.6$ ($F_y$ phase).
High symmetry points $\Gamma=(0,0)$, $X=(\pi,0)$, $M=(\pi,\pi)$ and
$Y=(0,\pi)$ in the 2D Brillouin zone are defined in the inset.}
\label{fig:sq2}
\end{figure}

Let us exemplify this in a situation where all couplings are AF,
and $J_z>J_x$, see Fig.~\ref{fig:sq2}. The large peak of $S^z(\vec q)$
found at $ M=(\pi,\pi)$ for $I=0.3J_c$ indicates that the corresponding ground
state is of the $G_z$ type, that is N\'eel-ordered with spins along the $z$ axis.
We encountered similar ordering features for $I/J_c$ varying from very small
to very large values, keeping the compass amplitudes fixed such that
$0<\phi<\pi/4$.
In the $I \gg J_c$ limit, this can be interpreted as follows:
small compass couplings break the $SU(2)$ symmetry of Heisenberg
interactions and select an easy axis for the N\'eel order; it is obvious
here that this
easy axis is $z$. In the limit $I \ll J_c$, the selection of $G_z$ order
is to be understood differently: we recall that compass couplings alone,
on a $L_x \times L_z$ cluster, select a class of $2^{L_x}$
\textit{columnar states}, characterized by long-range N\'eel-type
correlations along columns but short-range correlations along rows.
These states are separated from higher energy levels by a large gap
which is $\simeq 4J_z$ in the Ising limit $J_x/J_z \rightarrow 0$.

\begin{figure}[t!]
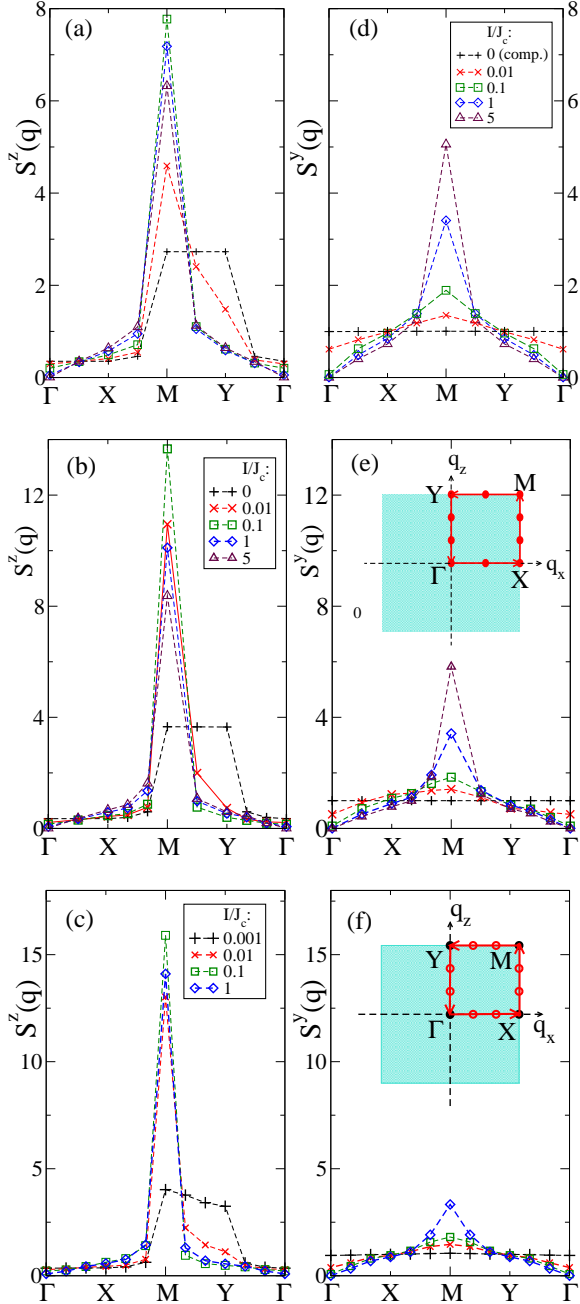

\begin{center}
\includegraphics[width=7.6cm]{SQyz_16t5.eps}\\
\vspace{0.4cm}
\includegraphics[width=7.6cm]{SQyz_24t5xk.eps}\\
\vspace{0.4cm}
\includegraphics[width=7.6cm]{SQyz_36t5.eps}
\caption{(Color online)
Structure factors (a-c) $S^z(q)$ and (d-f) $S^y(q)$ obtained for the
isotropic compass model $J_x=J_z$ ($\phi=\pi/4$) and increasing values
of the Heisenberg interaction $I$ for clusters of different size:
(a,d) $N=16$, (b,e) $N=24$, and (c,f) $N=36$ clusters.
The $N=24$ cluster is rectangular with $(L_x,L_z)=(4,6)$, as indicated
by the points in (b,e).
The high symmetry points $\Gamma$, $X$, $Y$ and $M$ are as in Fig.
\ref{fig:sq2}. The compass quasi-1D order is characterized
by constant values of $S^z(q)$ along the $M-Y$ line. The onset of the
AF order in the TL is evidenced by the increasing maximum of
$S^z(q)$ at $M=(\pi,\pi)$ which is visible already for very small
values of $I>0$.
}
\label{fig:sq1}
\end{center}
\end{figure}

This semi-ordered nature of the GS is reflected in
Fig.~\ref{fig:sq2}(a) for $I/J_c=10^{-6}$
by a structure factor spread over all momenta of the form
$\vec q = (q_x,\pi)$ --- for $J_x=I=0$, one would indeed have
$S^z(\vec q)=L_z \delta_{q_z,\pi}$. The effect of small $J_x$
compass couplings is mostly to reduce slightly the difference of
$S^z(\vec q)$ at $(q_x,\pi)$ and $(q_x,q_z \ne \pi)$, respectively.
In contrast, even very small Heisenberg couplings have a much stronger
effect, seen in the example of Fig.~\ref{fig:sq2}: they result in a strong
enhancement of $S^z(\pi,\pi)$, compared to $S^z(q_x,\pi)$ for $q_x\ne \pi$.
The fact that this enhancement is much stronger in this case for
$I/J_c=10^{-6}$, than in the anisotropic case for $I/J_c=10^{-3}$ of
Fig.~\ref{fig:sq1}(c), can be understood within the effective pseudospin model
which we will describe in Sec.~\ref{sec:eff}. To explain this, we notice that
the $2^{L_x}$ columnar states include two N\'eel-like states,
where spins on each $x$-oriented bond connecting neighboring columns are
antiferromagnetically arranged. These states are favored
over the other $2^{L_x}-2$ columnar states
by AF Heisenberg couplings, with arbitrarily small amplitude $I$ ---
namely by the $\sigma^z \sigma^z$ components of these
couplings on horizontal bonds (transverse components
of Heisenberg couplings have no matrix element between distinct
columnar states).
These favor an AF arrangement of the $z$ component of spins on
$x$-oriented nearest neighbor bonds, which at the global scale result
in the $G_z$ order.
In Sec. \ref{sec:eff} we provide an explanation, based on the
formalism of pseudospins $\vec{\tau}_j$,
for the fact that such an ordered phase can be selected in the TL, even for
infinitesimal Heisenberg couplings.
When these become larger, the $G_z$ order remains the most favorable,
and is almost fluctuation-free for $I \lesssim J_z$ (evidenced in
Fig.~\ref{fig:sq2}(a), for $I/J_c=0.3$, by $S^z(M)/N \simeq 0.9$
close to the maximal allowed value $1$).

A similar situation is also found for negative, moderate Heisenberg
couplings $I<0$, as shown in Fig.~\ref{fig:sq2}(b) for $I/J_c=-0.3$.
Here, among columnar states, those favored by Heisenberg
couplings have FM correlations on $x$-oriented bonds. The selected
ordered patterns define now a $C'_z$ ordered phase (see inset in the
corresponding region of the phase diagram for the AF CH model displayed
in Sec.~\ref{sec:phd}) which manifests itself by a peak at $\vec q=Y$ in
the structure factor $S^z(\vec q)$.

With the same value of anisotropy parameter $\phi$, but large negative
Heisenberg couplings ($0< J_c \ll -I$), the selected order follows from
a different mechanism. The dominant Heisenberg interactions tend to
favor a FM phase, with possibly an easy axis due to the
anisotropy induced by compass couplings. Actually, one
sees in Fig.~\ref{fig:sq2}(b) that $S^y(\vec q)$
has a distinct maximum for $I\simeq -0.6J_c$
at $\vec q=\Gamma$, while $S^z(\vec q)$ (and $S^x(\vec q)$, not shown)
do not display such a peak. This indicates that the FM order develops
already for moderate values $I=-0.6J_c$ when the Heisenberg coupling
strength $|I|$ increases, and $y$ axis is selected as the easy axis in
spin space.
Here, unlike in previous cases, a FM-ordered pattern cannot correspond
to any of the columnar states favored by the compass interactions alone
(since these would lead to AF correlations on either $x$ or $z$ bonds),
and the FM order along the $y$ axis appears as a compromise, which can be
understood as follows:
The AF compass couplings frustrate the dominant FM Heisenberg ones; but
since they act only on two spin components, the system avoids this
frustration by rotating spins away from the $xz$ plane to the $y$ axis
where they fully profit from Heisenberg interactions.
Although this mechanism is based on a classical picture, it explains the
behavior observed over a wide range of $I/J_z$ including the example
shown $I=-0.6J_z$ where frustrating compass terms are of amplitudes
comparable to Heisenberg ones. Several other ordered phases,
displayed in the phase diagrams shown in Sec. \ref{sec:phd},
are found when at least one of the coupling constants $\{J_x,J_z\}$ is
negative. For small $|I|/J_c$ they result always from the selection
of a pair of compass states by small Heisenberg couplings, because
they couple components along the easy axis ($z$ or $x$) of
spins neighboring on $x$ or $z$ bonds, respectively.

We turn now to the case where compass couplings are isotropic, i.e.,
$J_z=J_x$. There, in absence of Heisenberg couplings the low-energy
states of a $L_x \times L_z$ system consist not only of $2^{L_x}$
\textit{columnar states}, but also of $2^{L_z}$ \textit{row-type states}
The latter ones have spins oriented along $x$ and long-range correlated
along rows, but not along columns. As shown in Ref.~\onlinecite{Dor05}
this situation corresponds to a first order transition point between two
distinct phases characterized by either column- ($J_z>J_x$) or row-
($J_z<J_x$) -type ground states in the TL. Yet, as in the previously
discussed anisotropic case, in the isotropic one $J_x=J_z$ small AF
Heisenberg couplings select among these states only a small number, here
four. The selected states are here N\'eel states: two of them have spins
along $z$, while the two others with spins along $x$ are selected
within the class of row-type states. These four states are {\it a priori}
characteristic of a $\mathbb{Z}_2 \times \mathbb{Z}_2$ ordered phase,
breaking spontaneously not only translation symmetry, but also the
symmetry $\left\{\sigma^x_{\vec r},\sigma^y_{\vec r},
\sigma^z_{\vec r}\right\}\rightarrow\left\{\sigma^z_{R(\vec r)},
-\sigma^y_{R(\vec r)}, \sigma^x_{R(\vec r)}\right\}$,
where $R$ is the reflection w.r.t. the $z=x$ diagonal in real space.

The $\mathbb{Z}_2 \times \mathbb{Z}_2$ ordered phase is characterized by two
peaks in structure factors
$S^x(\vec q)$ and $S^z(\vec q)$, both at $\vec q = M$. Structure factors
$S^z(\vec q)$ are shown in Figs.~\ref{fig:sq1}(a)-\ref{fig:sq1}(c) for 3
different clusters of $N=16,24,36$ sites and several values of $I/J_c$
ranging from $0$ to $5$. In contrast with the anisotropic case where
such a peak can have a value close to the maximum allowed ($N$ being the 
number of sites), here peak amplitudes are
limited  by sum rules to $N/2$ on isotropic clusters with $L_x=L_z$ [Note that
if the cluster is anisotropic the peak amplitudes can slightly exceed the value
$N/2$, as in Fig.~\ref{fig:sq1}(b) for $(L_x,L_z)=(4,6)$ and $I/J_c=0.1$.]
A more striking feature is that these peaks grow very fast with small $I$: peak
amplitudes exceeding $75\%$ of the maximal value are attained for $I=0.01J_c$
on the largest clusters. In a situation with a slight compass anisotropy
$|J_x-J_z| \ll J_z$ one can show (see Ref. \onlinecite{Tro10}) that
the order develops as soon as $I \gg J_{\rm col}$ with $J_{\rm col}$ vanishing
exponentially with increasing $L_z$; this argument extends to the isotropic
case (see also Sec.\ref{sec:eff} for details). In contrast, we also show
structure factors $S^y(\vec q)$ in Figs. \ref{fig:sq1}(d)-\ref{fig:sq1}(f).
A peak is also observed when Heisenberg and compass couplings are of
similar values,
but the peak amplitude is almost independent of system size (consider
e.g. the $I=J_c$ case). For these reasons one can conclude that the
N\'eel order, with spin directions $x$ and $z$ equally favored over the
$y$ direction, is selected in the TL for arbitrarily small Heisenberg
couplings, in the whole range of values $I/J_c>0$ in the isotropic AF
case. We will see in the next Section that this order is unstable even
to infinitesimal variations of $J_z-J_x$ --- depending on the sign of
this quantity, either the N\'eel patterns with spin along $z$ or those
with spins along $x$ are favored, and one recovers the $G_z$ or $G_x$
phases.

\section{Phase diagram}
\label{sec:phd}

The CH model reveals a large variety of ordered ground states as
function of the interaction parameters $J_x/J_z$ and $I/J_c$. Some of
these phases were described in the previous Section. The determination
of the ground state phase diagram, and the characterization of QPTs
(as, for instance, the $G_z-G_x$ transition discussed above) will be the
object of the present Section. We will first give a classification of
the possible phases expected in the classical limit of the model; then
we will determine the phase diagram, first restricting ourselves to the
AF CH model (case $J_z>0$), before addressing
properties specific to the FM CH model (case $J_z<0$).

\subsection{Ordered phases of the CH-model}
\label{phases}

To analyze the phase diagram of the CH model, we
consider first the classical (or large $S$) limit, where one regards
spins as vectors living on a unit sphere. This
is of prime interest, since we will see that all ordered phases of the model
are found in this limit. To draw a tentative classical phase diagram one
needs to compare the ground state energies $E_0$ associated to these
different phases.
In Table~\ref{tabphi} we present a list the candidate phases $\Phi_\alpha$.
For each phase the index $\alpha \in{x,y,z}$ denotes
the easy axis or spin direction favored, while the capital letter in
$\Phi_\alpha$ indicates the type of spatial structure or correlation
pattern, i.e., $G$ for N\'eel-type AF phase, $F$ for FM phase, and $C$
for columnar or $C$-type AF order, i.e., with nearest neighbor spin
correlations being AF for one bond direction and FM for
the other. By convention, the presence of a prime in $\Phi'$ for
$C$-type phases, as e.g. for $C'_z$, indicates that nearest neighbor
correlations are AF on bonds
where compass interactions couple spin components along the easy axis.
Furthermore for each phase we indicate the momentum $\vec q$ such that
$S^\alpha(\vec q)/N$ stays finite in the TL.
For instance, the $F_z$ phase is the FM phase with spins along the
$z$ axis --- this order implies that the structure factor $S^z(\vec q)$
develops a peak at $\vec q= \Gamma$ of finite amplitude in the TL.

\begin{table}[b!]
\caption{ Classification of ordered phases of the CH model. For each
phase $\Phi$, the easy spin axis $\alpha$, the ordering wave vector
$\vec q$ and the classical energy per site $E_0(\Phi)$ are given.
Here $\Phi$ indicates either a FM ($F$), $G$-type AF ($G$), or
$C$-type AF ($C$) phase. }
\vskip .2cm
\begin{ruledtabular}
\begin{tabular} {c  c  c  c c c  c  c  c  c }
$\Phi$ \hst & $\vec q$ \hst & $\alpha$ \hst & $E_0(\Phi)$ \hst  & &  & $\Phi$
\hst & $\vec q$ \hst & $\alpha$\hst & $E_0(\Phi) $ \hst \\
\hline
$F_z$    \hst & $\Gamma$   \hst  &  $z$  \hst  &  $J_z+2I$ \hst & & & $G_z$
\hst &M  \hst &  $z$   \hst & $-J_z-2I$ \hst\\
$F_x$  \hst & $\Gamma$  \hst &  $x$  \hst  & $J_x+2I$ \hst & &
&  $G_x$ \hst & M  \hst &  $x$ \hst
& $-J_x-2I$ \hst \\
$F_y$ \hst & $\Gamma$ \hst &  $y$  \hst
  & +$2I$ \hst & & & $G_y$ \hst & M  \hst &  $y$  \hst & $-2I$   \hst \\
$C_z$  \hst & X   \hst &  $z$    \hst
   & +$J_z$ \hst & & &  $C_x$ \hst & Y  \hst &  $x$   \hst & +$J_x$ \hst \\
$C'_z$ \hst & Y \hst  &  $z$    \hst
   & $-J_z$\hst & & & $C'_x$ \hst  & X \hst &  $x$   \hst & $-J_x$  \hst \\
\end{tabular}
\end{ruledtabular}
\label{tabphi}
\end{table}

Eventually we give the ground state energy (per site) $E_0(\Phi_\alpha)$ of
each phase in the classical limit. 
The classical energy per site depends linearly on compass coupling amplitudes
$J_x$ and $J_z$, and on Heisenberg amplitude $I$, but only one or two of these
amplitudes appear(s) in the expression of $E_0(\Phi_\alpha)$ for a given phase.
One can consider for instance 
the $C_x$ phase, which has spins along $x$, and the ordering wave vector 
(at which $S^x(\vec q)/N$ is finite in the TL) $Y=(0,\pi)$, i.e., spin
correlations are FM along $x$ bonds and AF along $z$ bonds. 
In the classical limit only compass couplings contribute to its energy
per site $J_x$, since the contributions of Heisenberg couplings on $x$ bonds
and on $z$ bonds cancel each other.

It is clear that each of these phases is stabilized, at least in the classical
version of the model, when the coupling amplitude(s) entering $E_0(\Phi)$
is/are much larger in absolute value than other amplitude(s) --- for phases
$F_{x/z}$ and $G_{x/z}$ the condition is that both (compass and Heisenberg)
amplitudes involved in $E_0(\Phi)$ have equal signs and that their sum
is much larger than the amplitude of any frustrated interaction.
The determination of the domains of stability of these
phases, first in the classical limit and then in the $S=1/2$ model,
will be described hereafter, focusing first on the AF case $J_z>0$.

\subsection{Phase diagram: \textit{antiferromagnetic case} $J_z>0$.}
\label{sec:phdAF}

\subsubsection{Classical approach and symmetry relations}
\label{sec:sym}

\begin{figure}[t!]
\includegraphics[width=8cm]{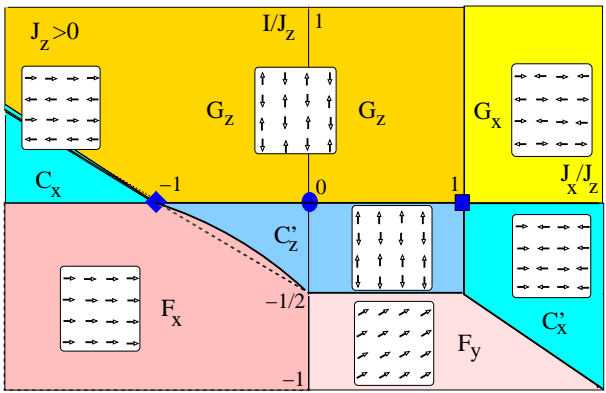}\\
\caption{(Color online) Phase diagram of the Compass-Heisenberg model
in the $(J_x,I)$-plane for fixed AF
interaction $J_z=1$. Long-range order is stabilized by any
finite $I$. Square ($J_x=J_z$, $I=0$) and diamond ($J_x=-J_z$, $I=0$)
at the compass line
indicate multi-critical points, where in each case four ordered phases meet.
The spin order of the different phases $\{G_z, G_x,C'_z, C_x, C_x',F_x, F_y\}$
is depicted in a corresponding inset, and the subscript $\alpha=x,y,z$
indicates the type of symmetry breaking in spin space, see Table I.
The QPTs between $F_x$ and $C'_z$ phases, and between $C_x$ and $G_z$ phases
(solid lines) are modified by quantum corrections w.r.t. the
corresponding classical transitions (dashed straight lines).
}
\label{fig:phdAF}
\end{figure}

Here we consider the case $J_z>0$, where most of the phases listed in
Table~\ref{tabphi} --- actually all but $F_z$, $C_z$ and $G_y$ can be
stabilized
depending on the values of interaction parameters $I/J_z$ and $J_x/J_z$.
By using the classical energies as given in this Table and determining,
for fixed $I/J_z$ and $J_x/J_z$, which of these energies is the lowest one
(with the assumption, which will appear as justified in the following,
that no other phase is stabilized in a finite volume
of the phase space determined by these two parameters), we
find the classical phase diagram represented in Fig.~\ref{fig:phdAF},
with transitions between two phases indicated by dashed straight lines
(coinciding in some cases with continuous lines).
The classical phase boundaries are straight lines in the
present parametrization, because the
classical energies depend linearly on the various coupling amplitudes. Not 
surprisingly, for all interactions being AF ($J_x, J_z, I>0$),
N\'eel order is always favored, with a
$G_z$ or $G_x$ phase depending on the sign of $J_z-J_x$. More interesting is
the extent of the $C'_z$ phase for moderate, negative $I$. Although in this
phase, due to the $z$ orientation of spins, the $J_x$ compass couplings are
frustrated,
their sign matters for the stability of this phase: its extent in
the phase diagram is smaller for $J_x<0$ (there it competes with the $F_x$
phase stabilized by $J_x$ couplings) than for $J_x>0$. In the latter case
it competes with the $F_y$ phase where $J_x$
couplings are inactive, which explains why the $C'_z-F_y$ transition line
is independent of $J_x/J_z$.

Among these phase transitions, several ones occur on transition lines
in the phase diagram of Fig.~\ref{fig:phdAF} which follow from symmetry
considerations, and are thus insensitive to quantum fluctuations. The
simplest example is that of the $G_x\leftrightarrow G_z$ transition:
intuitively, one
can guess that it can occur only for $J_z=J_x$ (and $I>0$), but one can
also notice that the transformation defined by:
\beq
(\tau_{\vec s}^{'x},\tau_{\vec s}^{'y},\tau_{\vec s}^{'z}) = (\sigma_{\vec
  r}^z,\sigma_{\vec r}^y,-\sigma_{\vec r}^x) ,
\eeq
where $\vec s = R(\vec r)$ and $R$ is a spatial rotation of $\pi/2$ around a
reference site, allows us to rewrite the CH Hamiltonian as follows:
\bea
H&=& J_x \sum_{i,j} \tau_{i,j}^{'z} \tau_{i+1,j}^{'z}
+ J_z \sum_{i,j} \tau_{i,j}^{'x}\tau_{i,j+1}^{'x} \nn\\
&+&I \sum_{i,j} \vec{\tau'}_{i,j} \cdot
\left(\vec{\tau'}_{i,j+1} + \vec{\tau'}_{i+1,j}\right).
\eea
In other words, this transformation maps the domain $J_z<J_x$ of the
phase diagram onto the domain $J_z>J_x$ and vice-versa, and if a point
with $J_z<J_x$ is in the $G_z$ phase it implies that its image by this
transformation is in the $G_x$ phase. Only at the transition line
$J_x=J_z$ the CH Hamiltonian is invariant
when this transformation is applied, which means that
the $G_x\leftrightarrow G_z$ transition has to occur there
(unless another intermediate
phase is stabilized, which is not expected). For $I<0<J_x,J_z$, the same
transformation is a bijection between each point of the $C'_z$ phase,
with given $I/J_z$ and $\phi=\arctan(J_x/J_z)$, and a point in the
$C'_x$ phase, with the same value of $I/J_z$ and anisotropy
parameter $\pi/2-\phi$.
This implies that the transition line between these two phases is fixed
to the line $J_x=J_z$ as in the classical limit, under the condition
that the Heisenberg amplitude $|I|$ is too small to stabilize the FM
$F_y$ phase which would be favorable otherwise.
One can notice --- here at the classical level but this feature is actually
conserved in the quantum model --- that the isotropic point of the compass
model, with $J_x=J_z>0$ and $I=0$, is unstable to even infinitesimal
variations of either $J_x-J_z$ or $I$; depending on the sign of both
quantities four different phases can be selected, so that this point can be
seen as a quadricritical
point in the context of the CH model (we do not mean
by this that correlations are algebraic there, but simply that four phases
meet at this point).

Another transition characterized by an additional symmetry is the one
between the $C'_z$ and the $F_y$ phases, obtained by varying the
Heisenberg amplitude and keeping $\phi \in (0;\pi/4)$ fixed, and
stabilized for $0<-I \ll J_c$ and for $-I \gg J_c$, respectively.
In this case one can make use of a transformation defined by:
\beq
\{\tau_{i,j}^{''x},\tau_{i,j}^{''y},\tau_{i,j}^{''z}\} = \{(-1)^i
\sigma_{i,j}^x,\sigma_{i,j}^y,(-1)^i \sigma_{i,j}^z\} .
\label{tczfy}
\eeq
Reexpressing all couplings in function of $\vec{\tau'}$ operators, one obtains
the following Hamiltonian:
\bea
H &=& H_x[{\tau^{''x}}] \nonumber \\
&+& \sum_{i,j} \left\{ I \left( \tau_{i,j}^{''y} \tau_{i+1,j}^{''y} +
\tau_{i,j}^{''y} \tau_{i,j+1}^{''y}\right) \right. \nn \\
&-&\left. (I+J_z) \tau_{i,j}^{''z}\tau_{i+1,j}^{''z} + I \tau_{i,j}^{''z}
\tau_{i,j+1}^{''z} \right\}.
\label{eq:taucf}
\eea
Here $H_x[{\tau^{''x}}]$ is a function of only $x$-components of $\vec{\tau''}$
spins, necessarily invariant under further rotations
of spins along the $x$
axis (not combined here with any spatial symmetry but the identity).
From the expression of $H-H_x[{\tau^{''x}}]$ in Eq.~({\ref{eq:taucf}}),
one sees that for $I=-J_z/2$ such rotations leave $H$ invariant:
there, an extra $U(1)$ symmetry appears, so that all states with fully
polarized $\vec{\tau''}$ spins in the $yz$ plane are degenerate ground states.
These contain the two degenerate ground states of the $F_y$ phase,
as well as those of the $C'_z$
phase (which is also FM in terms of $\vec{\tau''}$ spins):
necessarily, the transition between both phases has to occur there. Notice
that, unlike for previously discussed $G_x-G_z$ and $C'_x-C'_z$ transitions,
here no mapping from the $F_y$ to the $C'_z$ phase is allowed away from the
transition line.

In contrast, another phase transition occurring in the classical phase
diagram found in Fig.~\ref{fig:phdAF} for $I<0$ and $-J_z<J_x<0$,
namely the one between $C'_z$ and $F_x$ phases, is not fixed by any
symmetry relation. Not only one cannot find a mapping between both phases
as in the $G_z\leftrightarrow G_x$ case, but also if one looks for a 
transformation of the type given by Eq.~(\ref{tczfy}), there is no extra 
symmetry ($U(1)$ or other) at the classical transition line $I=-(J_z+J_x)/2$.
In consequence, the corresponding transition line can be shifted by
quantum fluctuations --- which of the two phases is stabilized by those
fluctuations at the $I=-(J_z+J_x)/2$ line is
one of the questions addressed in the next paragraph.

\subsubsection{Phase transitions in the quantum CH model}
\label{sec:pt}

\begin{figure}[t!]
\includegraphics[width=7.5cm]{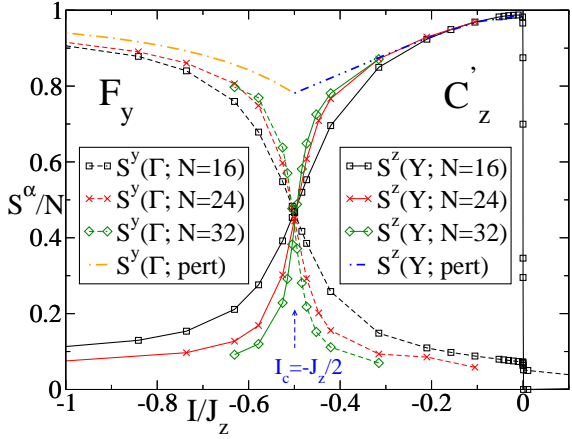}\\
\caption{\label{fig:tr1} (Color online)
Evolution of order parameters $S^y(0,0)/N$ and $S^z(0,\pi)/N$
across the $C'_z\leftrightarrow F_y$
transition, by varying $I/J_z$; anisotropic compass interactions
have the same parameter $\phi=\pi/10$ as in Fig.~\ref{fig:sq2}.
The cluster $N=24$ has dimensions $L_x=4, L_z=6$.
The dashed-dotted lines correspond to perturbative estimations of order
parameters --- given for the $C'_z$ phase by Eq.~(\ref{pertsz}).
The transition coincides here with the classical value $I_c=J_z/2$.
}
\end{figure}

We turn now to the phase diagram of the $S=1/2$ CH model and investigate
those aspects which show up as a result of quantum fluctuations.
In Sec.~\ref{phases} the magnetically ordered phases
were selected either for $|I| \ll J_c$ or $|I| \gg J_c$ , and subsequently
their respective stability in the classical phase diagram was discussed.
There we made the implicit assumption that no other phase occurs in
an intermediate range of $I$;
we will now see that this assumption is justified even in the quantum model.

\begin{figure}[t!]
\includegraphics[width=7.5cm]{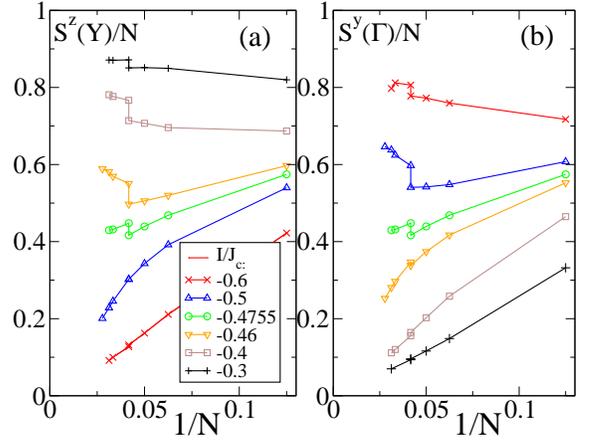}\\
\caption{\label{scal} (Color online)
Size-scaling of the order parameters of (a) $C'_z$ and (b) $F_y$ phases for
fixed anisotropy parameter $\phi=\pi/10$ and for values of $I/J_c$ at and
close to the value $I=-0.4755J_c$ of the transition between both phases.
The data points are obtained with the $N=8,16,20,24,30,32,36$ clusters.
Two data points for $N=24$ correspond to two different rectangular
clusters: $4\times 6$ and $6\times 4$.}
\end{figure}

The first case to be addressed is the transition between the $C'_z$ and
$F_y$ phases, occurring by increasing $|I|/J_z$ from $0$ to infinity,
with $I<0<J_z$ and fixed $J_x/J_z \in (0;1)$.
The classical approach (see Sec.~\ref{sec:sym}) predicts a transition
between those phases at $I=-J_z/2$. In the case of the quantum model we
study in Fig.~\ref{fig:tr1}
the evolution of the spin structure factors $S^z(Y)$ and $S^y(\Gamma)$
corresponding to the $C'_z$ and $F_y$ phases as function of $I/J_z$.
The data in Fig.~\ref{fig:tr1} calculated at fixed $\phi=\pi/10$
indicates that no intermediate phase is stabilized in a finite range of
$I$ between the $C'_z$ and $F_y$ phases, since on both sides of the
classical transition point $I_c=-J_z/2$ $(I_c\simeq -0.4755J_c)$,
either $S^z(Y)/N$ or $S^y(\Gamma)/N$ takes large values.
This is a clear evidence of long-range magnetic order of either $C'_z$
or $F_y$-type, respectively. The transition is clearly detectable
already on clusters of moderate size, by the sharp evolution in its vicinity
of structure factors as function of $I/J_z$: the maximal slopes are found
\textit{exactly} at $I=I_c$. The size scalings of both order parameters
(i.e., of the related structure factors divided by $N$) are shown in
Fig.~\ref{scal} and provide evidence of this transition on a more
quantitative level, with each order parameter exhibiting a clear
change of behavior at $I=I_c$: for $|I|<|I_c|$ the scalings indicate
that $S^z(Y)/N$ and $S^y(\Gamma)/N$ take, respectively, a finite
value or 0 in the TL, while for $|I|>|I_c|$ it is the contrary ---
eventually in the particular case $I=I_c$ both order parameters scale
down to a common finite value, confirming as well that this transition
point can also be seen as an intermediate phase where the U(1) symmetry
is spontaneously broken and the ground state is an XY-type ferromagnet
in terms of $\vec{\tau}^{''}$ spins defined in Eq.~(\ref{eq:taucf}).
The corresponding order parameter, $(S^y(\Gamma)+S^z(Y))/N$, would be
equal to 1 in the TL if terms $H_x[\tau^{''}_x]$ were absent from
Eq.~(\ref{eq:taucf}); in their presence,
for $\phi=\pi/10$, Fig.~\ref{scal} indicates that this order parameter
attains $\simeq 0.7(1)$ in the TL. Yet this XY-type order is sensitive
to infinitesimal variations of $I$, which
lower the symmetry of the Hamiltonian from $U(1)$ to $\mathbb{Z}_2$ and select
an easy axis $y$ or $z$ for the ordered phase.

Sufficiently far away from the above phase transition, quantum
corrections to the values
of order parameters (which would show, in the classical limit, a discontinuity
at $I_c$ with a jump/drop from $1$ to $0$ or from $0$ to $1$) are relatively
well estimated using second-order perturbation theory.\cite{Sch09}
There, the unperturbed Hamiltonian consists for a given phase of
components along the easy axis (e.g. $z$ in the $C'_z$ phase) of both compass
and Heisenberg couplings, while transverse components of these
couplings are regarded as perturbations.
One can illustrate this in the case of the $C'_z$ phase: the effects of
quantum fluctuations is to reduce (in absolute value) the correlation
between $\sigma^z$ components of spins, and thus the $C'_z$ order parameter.
For sites $\vec r \equiv (i,j)$ and $\vec s \equiv(i',j')$
situated at distance $d>1$ from each other, one obtains:
\begin{eqnarray}
\langle \sigma^z_{\vec r} \sigma^z_{\vec s} \rangle &\simeq& (-1)^{i-i'}
\left\{ 1-\frac{J_x^2}{4(2J_z+I)^2}-\frac{I^2}{(J_z-I)^2}
\right\} \nonumber \\
&\times&
\left\{1+\frac{J_x^2}{4(2J_z+I)^2}+\frac{I^2}{(J_z-I)^2} \right\}^{-1} .
\label{pertsz}
\end{eqnarray}
The prefactor $(-1)^{i-i'}$ cancels out with the
phase factor in Eq.~(\ref{saq}) giving the order parameter $S^z(Y)/N$,
which in the TL is equal to the absolute value of the expression in
Eq.~(\ref{pertsz}).\cite{noteco}
The perturbative estimate $S^z(Y;{\rm pert})$ of the
$C'_z$ structure factor is shown in Fig.~\ref{fig:tr1} for $\phi=\pi/10$,
and gives good agreement with the numerics, away from the transition to
the $F_y$ phase (here, for $|I|\lesssim 0.3$).
Similar estimates can also be obtained for other ordered phases,
like $S^y(\Gamma;{\rm pert})$, also shown in Fig.~\ref{fig:tr1}, in the $F_y$
phase: here the unperturbed Hamiltonian consists of couplings
$\sigma^y_{\vec r} \sigma^y_{\vec s}$ in $\cal{H}$.
Concerning absolute values, the agreement with numerics is less
accurate than in the $C'_z$ phase but the dependence on $I/J_z$
is correctly reproduced by the perturbative result.

\begin{figure}[t!]
\begin{center}
\includegraphics[width=7.5cm]{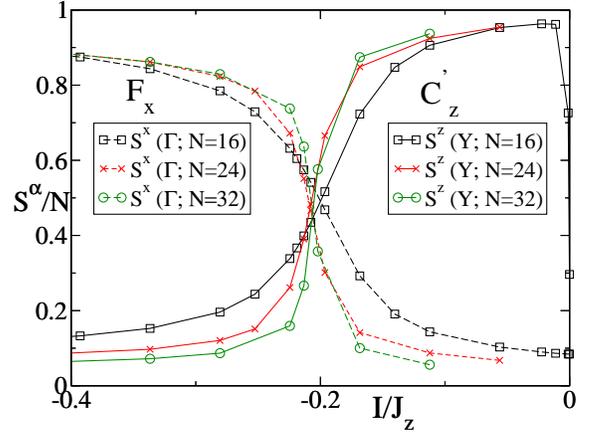}\\
\caption{\label{fig:tr2} (Color online)
Variation of order parameters $S^x(\Gamma)/N$ and $S^z(Y)/N$ across 
the $C'_z\leftrightarrow F_x$ transition as function of $I/J_z$
at fixed $\phi=-3\pi/20$, and for system sizes $N=16, 4\times6, 32$.
Here the quantum phase transition is shifted by quantum fluctuations
from $I_c^0/J_z=-0.245$ to $I_c/J_z=-0.21(1)$. }
\end{center}
\end{figure}

The spin structure factors are even more useful to study phase transitions
not characterized by additional symmetries, such that the phase
boundaries can be modified by quantum fluctuations. As an example
we focus on the $C'_z\leftrightarrow F_x$ transition: the relevant order
parameters are $S^z(Y)/N$ and $S^x(\Gamma)/N$.
We show in Fig.~\ref{fig:tr2} their
evolution as function of $I/J_z$, again for fixed $J_x/J_z$. For each
cluster size, the two curves have maximal slope at the same value of
$I/J_z$, which can thus be considered as a finite-size transition point.
But in contrast to the $C'_z\leftrightarrow F_y$ case, here this
transition point is cluster-dependent and distinct from the classical
one ($I_c^0 \simeq -0.245 J_z$ for $\phi=-3\pi/20$). The
dependence on $N$ of this finite-size transition point is rather weak, and
this allows us to locate approximately the transition point in the TL ---
for parameters of Fig.~\ref{fig:tr2}, it occurs at
$I_c/J_z \simeq -0.21(1)$.

The deviation of the latter transition point from the classical value
$I_c^0$ can be well estimated by evaluating energies of both phases
using second-order perturbation theory.
This approach gives the following estimates for the energies per site
of the two phases involved:
\bea
\label{eq:e2fx1}
E(F_x)\!&=& E_0(F_x) + \frac{J_z^2}{8J_x+12I}, \\
\label{eq:e2fxcz}
E(C'_z)\!&=& E_0(C'_z) - \frac{J_x^2}{8J_z+4I} - \frac{I^2}{J_z-I}.
\eea
Within this approach, the transition point is given by the value of $I$
for which $E(F_x)=E(C'_z)$: still in the case $\phi=-3\pi/20$, this
value is $I_c^{(2)}\simeq -0.2040 J_z$,
that is very close to $I_c$
estimated from order parameters of both phases. More generally, for
variable $J_x/J_z \in(-1,0)$, the transition line between $C'_z$ and
$F_x$ phases as estimated from Eqs. (\ref{eq:e2fx1}) and
(\ref{eq:e2fxcz}) is shown on the phase diagram, see
Fig.~\ref{fig:phdAF}. One sees there that the deviation from the
classical transition is always in the same manner, i.e., at the
classical line $I=I_c^0=-(J_z+J_x)/2$ quantum fluctuations
favor the $F_x$ phase at the expense of the $C'_z$ phase. This may
appear surprising since --- at least in unfrustrated Heisenberg models
--- a FM phase is an exact fluctuation-free ground state. In contrast,
orders with some AF bonds, like the $C'_z$ phase, are typically
accompanied by quantum fluctuations, increasing with the increasing
number of AF bonds.\cite{Rac02} The case of the $C'_z\leftrightarrow F_x$
transition is different: first, both phases are characterized by easy axes,
distinct from each other, and second, contributions from different bonds to
quantum fluctuations have to be considered separately. The contribution
of $z$-bonds to quantum fluctuations, thanks to the large amplitude
$J_z >|J_x|, |I|$ in the vicinity of the classical transition,
removes the degeneracy and stabilizes
the $F_x$ phase with respect to the $C'_z$ one.

\begin{figure}[t!]
\begin{center}
\includegraphics[width=7.7cm]{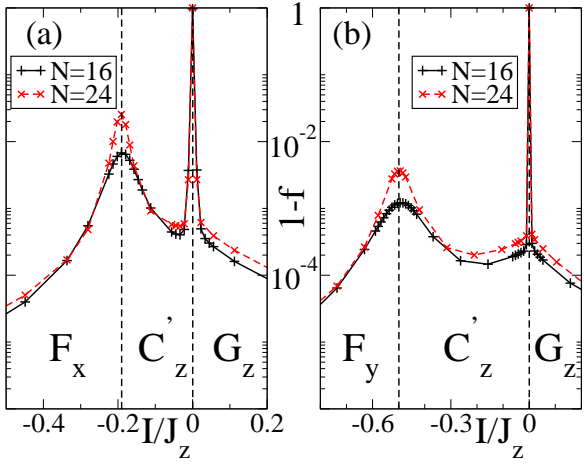}\\
\caption{\label{fig:fid} (Color online)
Evolution of $1-f$, where $f$ denotes the fidelity Eq. (\ref{defid}),
as function of $I/J_z$ for parameters (a) $\phi=-3\pi/20$ and (b) 
$\phi=\pi/10$,  and for two cluster sizes $N=16$ and $N=24$ (in 
both cases $L_z =4$).
Smooth peaks reflect phase transitions between the $F_x$ or $F_y$ phase
and the $C'_z$ phase that are continuous for finite systems. The sharp
peaks at $I=0$ indicate the $C'_z\leftrightarrow G_z$ transition.}
\end{center}
\end{figure}

Complementary to the detection from structure factors of the $C'_z$-$F_y$
and $C'_z$-$F_x$ phase transitions, one can also analyze the behavior as
function of $I/J_z$ of the fidelity,\cite{Wen07} defined as:
\beq
f(I) = |\langle \Psi_0(I+\delta I)|\Psi_0(I-\delta I)\rangle|,
\label{defid}
\eeq
where both ground states $|\Psi_0(I\pm\delta I)\rangle$
are computed for values of Heisenberg amplitudes differing
by $\pm \delta I$ ($\delta I=5\times 10^{-3}J_c$) from the nominal value
$I$. In Fig.~\ref{fig:fid} we plot the quantity $\ln(1-f)$ as a function of
$I/J_z$ for two clusters (with $N=16$ or $N=24$ sites), and for the two
values of $\phi$ corresponding to Figs.~\ref{fig:tr1} and \ref{fig:tr2},
respectively.
For a given cluster size and a given value of $\phi$, peaks are observed
on the $I/J_z$ axis at positions coinciding with the maximal slopes of
order parameters in Figs.~\ref{fig:tr1} and \ref{fig:tr2}. These peaks are
thus good indicators of phase transitions, here between the $C'_z$ and
other phases. Note that the peaks at $I=0$ (transition between $C'_z$ and
$G_z$ phases on the compass line of the phase diagram, either for $J_x>0$
or $J_x<0$) are much higher and thinner than those at transitions between
the $C'_z$ phase and either the $F_y$ or the $F_x$ phase, for $\phi>0$
and $\phi<0$, respectively. Indeed, the qualitative change in the ground
state occurs continuously at the $C'_z\leftrightarrow G_z$ transition,
but in a very narrow range of $I/J_z$ (estimated in Sec.~\ref{sec:eff}),
resulting in a sharp peak centered at $I=0$. In contrast,
at the transitions between the $C'_z$ and FM phases the peaks in
$\ln(1-f)$, centered around $I_c$ (up to small deviations resulting from
finite size), are much more smooth, characteristic of a continuous
transition. The latter behavior may be, {\it a priori}, an artefact
which follows from finite size,
while we have indications due to sharpening of peaks with increasing
system size that the transition may become first order in the TL.

\begin{figure}[t!]
\begin{center}
\includegraphics[width=7.7cm]{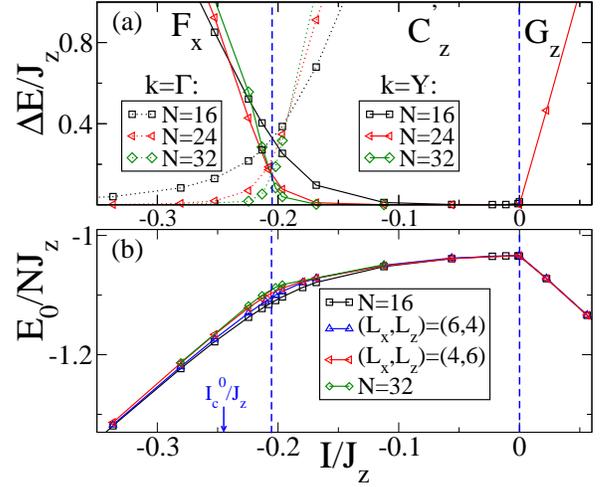}
\caption{\label{fig:eextr} (Color online)
(a) Lowest excitation energies $\Delta E$ with momentum either
  $k=\Gamma=(0,0)$ or $k=Y=(0,\pi)$,
across the $F_x-C'_z$ and $C'_z-G_z$ transitions, as function of $I/J_z$ at
$\phi=-3\pi/20$ and for different cluster sizes; (b) Ground state energy
$E_0$, found with momentum $\Gamma$,
reflecting the avoided crossing of the $F_x-C'_z$ transition at
$I/J_z=-0.21(1)$. }
\end{center}
\end{figure}

Eventually, the phase transitions in the CH model can also be addressed
by considering the low-energy spectrum, which we illustrate once again
on the example of the $C'_z-F_x$ transition for fixed $\phi=-3\pi/20$.
The dependence of the ground state energy per site $E_0$ on $I/J_z$,
shown in Fig.~\ref{fig:eextr}(b), is consistent with that of the fidelity:
$E_0/N$
for a given cluster varies smoothly in the vicinity of the transition,
but from the comparison between different cluster sizes it appears that
a cusp develops in the TL at $I=I_c$, which supports the picture of a
first order transition (in this limit) responsible for the peak of
$\ln(1-f)$ seen in Fig. \ref{fig:fid}.
The nature of the transition may also be examined by considering the
lowest excitation energies,
see Fig.~\ref{fig:eextr}(a). On each side of the transition, the lowest
excited state is found in a representation indicative of the symmetry of
the phase stable beyond the phase transition:

(i) In the $F_x$ phase, the ground state is found with
$\vec k = \Gamma$; but both states have opposite parity of
$\frac12 \sum_{\vec r}\sigma_{\vec r}^z$
(the latter quantity being conserved in the model).
The energy splitting between both states increases when approaching the
transition point, and for fixed $I/J_z$ decreases (exponentially, as far
as one can tell) to zero with increasing linear size. This means that
both states become degenerate ground states in the TL; among linear
combinations of them one finds states fully-polarized either along $+x$
or $-x$ in spin space, and which break spontaneously the global symmetry
under $\sigma_{\vec r}^x \rightarrow -\sigma_{\vec r}^x$.

(ii) In the $C'_z$ phase, the second lowest state, also with an
excitation energy decreasing rapidly to zero with increasing size, has
identical parity of $\frac12\sum_{\vec r} \sigma_{\vec r}^z$ as the ground
state; but a distinct momentum $Y=(0,\pi)$. Here these states, degenerate
in the TL, are characteristic of the $C'_z$-type order, with spontaneously
broken symmetry of translation by one lattice unit along $z$ bonds
--- or, in terms of symmetries in spin space, spontaneous breaking of
the global symmetry under the
$\sigma_{\vec r}^z \rightarrow -\sigma_{\vec r}^z$ transformation
takes place.

\begin{figure}[t!]
\begin{center}
\includegraphics[width=7.7cm]{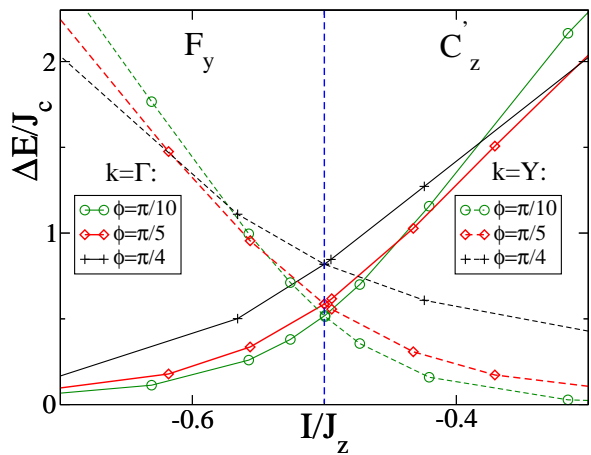}\\
\caption{\label{fig:eextr2}(Color online)
Lowest excitation energies $\Delta E$
for $N=16$ across the transitions between the $F_y$ and $C'_z$
phases (for three distinct values of $J_x/J_z$ between 0 and 1,
corresponding to $\phi=\pi/10$, $\pi/5$, and $\pi/4$).}
\end{center}
\end{figure}

At equal size, the positions of the crossings seen on Fig.~\ref{fig:eextr}
for the $C'_z\leftrightarrow F_x$ transition, and Fig.~\ref{fig:eextr2}
for the $C'_z\leftrightarrow F_y$ transition, match well the positions
of maximal slopes in
Figs.~\ref{fig:tr2} and \ref{fig:tr1}, respectively, and at the crossing
their common excitation energy seems to decrease to zero towards the TL.
Such level crossings on finite systems are thus good indicators of the
corresponding phase transitions.

Considering the various features of phase transitions described above,
we can distinguish several types of transitions. Those occurring at
$I\ne 0$,
such as between $C'_z$ and either $F_y$ or $F_x$ phases, require a
particular attention. Although no level crossing is observed in the
ground state on finite systems at these transition, several features
indicate that they might be of first order in the TL:
(i) the ground state energy, as function of the interaction parameter
driving the transition, seems to develop in the TL a cusp characteristic
of a first order transition; \cite{Ham86}
(ii) the ordered phases on each side of the transition
have distinct $\mathbb{Z}_2$
symmetry groups (or distinct spontaneously broken symmetries);
(iii) accordingly, a crossing occurs at the transition between the
two lowest excitations, found in different symmetry representations,
and each of these excitations becomes one of the two degenerate ground
states of the respective phase in the TL;
(iv) tentative scalings of order parameters suggest that
they jump at the transition between zero and a finite value in the TL.

However, these indications are no evidence yet for a first order character
at the transition as scalings may be biased by the small system sizes
available. More
importantly, the fact that the two competing phases have distinct
symmetries does not prohibit a continuous transition, although beyond
the Landau-Ginzburg paradigm, between these phases.\cite{Wen09} Here,
the vanishing of the lowest excitation energy can also signal that the
system becomes gapless at the transition. This is clear in the
$C'_z\leftrightarrow F_y$ transition along the $I=-J_z/2$ line: there,
the U(1) symmetry is spontaneously broken, and the finite values 
of $S^z(Y)$ and $S^y(\Gamma)$ in the TL are, in the rotated basis of
$\vec{\tau}^{''}$ spins,
the two components of the order parameter for an XY-type ferromagnet.
Schematically, by varying $I/J_z$ the ordered moment can be
rotated continuously from $z$ to $y$ at the transition point, in
contrast to typical first order transitions where hysteresis phenomena
usually occur. More generally, we will see in Sec.~\ref{sec:sw} that
each phase transition away from the $I=0$ line is characterized by the
vanishing of the anisotropy gap to spin waves, which is finite in the
$\mathbb{Z}_2$ ordered phases on each side of the transition.
In consequence, the hypothesis that these transitions are continuous not
only on finite systems but also in the TL is justified as well. We also
note that, whereas some of these transitions
(as the $C'_z\leftrightarrow F_y$ one) are particular, with an additional
symmetry at the transition line, the same features occur at transitions
not characterized by such additional symmetry.

Eventually we comment on the $I=0$ line of the phase diagram, which can
be seen as a transition line between distinct ordered phases (e.g.
between the $C'_z$ and $G_z$ phases by increasing $I$ from negative to
positive values). There, spin waves are gapped (except at isotropic
points where $|J_z|=|J_x|$); these transitions are not characterized by
the softening of spin waves, but rather of
column-flips introduced in Sec.~\ref{sec:chm} and which are gapless in
the TL for $I=0$. This transition line, where one recovers the compass model,
is characterized by the non-local invariants of Eq.~(\ref{piqj}), and the
evolution between two distinct $\mathbb{Z}_2$ ordered phases through
this line can hardly be classified as an usual first- or second-order
transition.

\subsection{Phase diagram in the \textit{ferromagnetic case} $J_z<0$}
\label{sec:fm}

\begin{figure}[t!]
\includegraphics[width=8cm]{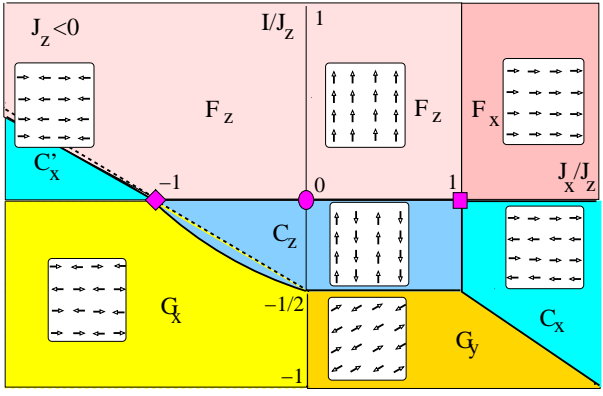}\\
\caption{\label{fig:phdF}(Color online)
Phase diagram of the compass-Heisenberg model in the
$(J_x/J_z,I/J_z)$-plane for fixed FM coupling $J_z=-1$.
Similar to Fig.~\ref{fig:phdAF}, square ($J_x=J_z$) and diamond
($J_x=-J_z$) indicate multi-critical points, and the spin order of the
different phases is depicted in insets. Phase labels follow the same
convention as in Fig.~\ref{fig:phdAF}.
The QPTs between $F_z$ and $C'_x$ phases, and between $C_z$ and $G_x$
phases (solid lines) are modified by quantum corrections w.r.t.
classical transitions (dashed straight lines).}
\end{figure}

In the previous paragraphs we restricted the analysis to the case $J_z>0$ and
described the corresponding phase diagram and phase transitions, by varying
two interaction parameters: $J_x/J_z$ and $I/J_z$. Here
we address the complementary case with FM couplings on $z$ bonds, i.e.,
$J_z<0$. The phase diagram of the {\it ferromagnetic} CH model is shown
in Fig.~\ref{fig:phdF}. It has many similarities with that of the AF CH
model in  Fig.~\ref{fig:phdAF}. There is an obvious difference in the
nature of the various phases, with e.g. for $J_x/J_z\in [0;1]$ and
$I/J_z>0$ a $F_z$ phase replacing the $G_z$ one of the $J_z>0$ case.
Another qualitative difference between both cases is that --- here we
assume $|J_z|>|J_x|$ --- the column-ordered states which are favored by
dominant compass interactions allow for
significantly less quantum fluctuations in the present $J_z<0$ case
than in the $J_z>0$ one. Concretely, this comes from the fact that
Heisenberg terms on vertical bonds are inactive on columns with all spins
aligned; this difference matters for the structure of the low-energy
spectrum. Eventually, another motivation to study the $J_z<0$ case
follows from a possible qualitative description of arrays of NV centers
coupled by quasi-short-ranged dipolar interactions
[see Eq.~(\ref{eq:Hdip})]: the situation most likely captured within the
CH model is with FM compass couplings $J_{x,z}<0$, accompanied by AF
Heisenberg couplings.

We address here the main ground state properties of the FM CH model by
considering first the classical limit. There, the phase diagram
has actually the same topology as in the AF case: transition lines have
the same positions and only the nature of the
long-range order in each individual phase, stable in a particular range
of $\{J_x/J_z,I/J_z\}$ parameters, is determined by the sign of $J_z$.
This is because \textit{in this limit} and due to the absence of
interactions between spins of the same sublattice (in terms of bipartite
sublattices), a transformation reversing the signs of all
couplings and simultaneously changing spins $\vec{\sigma}_{\vec r}$ to
$-\vec{\sigma}_{\vec r}$ on one sublattice leaves the energy
unchanged --- thus the ground states of both phases
considered are related by a spin reversal on one sublattice.

Coming back to the quantum model: here the shapes of the  phase diagrams
of the FM and AF model differ from each other in the same regions where
they differ from their respective classical counterparts, since the
transition lines which are not fixed for symmetry reasons are
differently affected by quantum fluctuations in the $J_z<0$ case than in
the $J_z>0$ case (Fig.~\ref{fig:phdAF}). The case to compare to the
previously discussed $C'_z\leftrightarrow F_x$ transition is here the
$C_z\leftrightarrow G_x$
transition, which classically occurs on the line $I=-(J_z+J_x)/2$ for
$0<J_x<-J_z$. Here as well, one can estimate the energies per site of
both phases in second order perturbation theory:
\begin{eqnarray}
E(C_z)\! &=&\! J_z - \frac{(2I+J_x)^2}{8|J_z|-4I},\\
E(G_x)\! &=&\!-J_x-2I- \frac{I^2}{3I+J_x} -\frac{(2I+J_z)^2}{12I+8J_x}.
\label{eq:e2czgx}
\end{eqnarray}
From this
one finds that on the line of the classical phase transition the energy
of the $C_z$ phase is lower due to quantum fluctuations than that of the
$G_x$ phase. This implies that, when taking quantum fluctuations into
account, the $C_z\leftrightarrow G_x$ transition line in Fig.~\ref{fig:phdF}
must have the opposite curvature to that of the $C'_z-F_x$ line in
Fig.~\ref{fig:phdAF}.
Here again, the perturbative estimation of transition points, which
yields for example $I_c/J_z=-0.295$ at $J_x/J_z=-0.5$, matches well with
the numerical estimates from the data of structure
factors (not shown) obtained with finite clusters.

\section{Spin wave excitations}
\label{sec:sw}

In the previous Sections we have mostly focused on ground state
properties of the CH model, and considered lowest excitation energies
merely as a tool to characterize the symmetry of ordered phases and to
locate phase transitions. Here we provide a description of the lowest
excitations characteristic of the various ordered phases in the TL,
these excitations are as usual spin waves; for this we will use linear
spin-wave (LSW) theory and see that this describes
efficiently the lowest single-magnon branches.

We begin the analysis of spin waves with the case of a FM phase, namely
the $F_z$ phase corresponding to $J_z, I<0$ and $|J_x|<|J_z|$. The
classical ground state with all spins pointing along $+z$ corresponds to
the vacuum of Holstein-Primakoff bosons $\{a^\dagger_{\vec r}\}$,
defined by the following transformation:
\begin{equation}
S^z_{\vec r}=S-a^\dagger_{\vec r} a_{\vec r}^{}=S-n_{\vec r}, \hskip .5cm
S^+_{\vec r} = \sqrt{2S}
\sqrt{1-\frac{a^\dagger_{\vec r} a_{\vec r}^{}}{2S}}\,a_{\vec r}.
\end{equation}
Here $\vec{S}_{\vec r}=\frac12\vec{\sigma}_{\vec r}$ are the usual
spin-$1/2$ operators. Due to the lack of SU(2)-invariance of
the model, after linearization the spin-wave Hamiltonian contains not
only $a^\dagger_{\vec r} a_{\vec s}$-type but also
$a_{\vec r} a_{\vec s}$-type terms, that do not conserve the number of
bosons:
\begin{eqnarray}
H_{\rm LSW} =4S \sum_{\vec r} \Big\{ \big[ I a^\dagger_{\vec r}
(a_{\vec r +\vec{e}_x}+a_{\vec r+\vec{e}_z}) + {\rm H.c.} \big]  \nn\\
  - (4I+2J_z) n_{\vec r} + \frac12 J_x
(a^\dagger_{\vec r} a_{\vec r + \vec{e}_x} +
a_{\vec r} a_{\vec r+\vec{e}_x} + H.c.) \Big\},
\label{eq:hlsw}
\end{eqnarray}
with $\vec{e}_x=(0,1)$ and $\vec{e}_z=(1,0)$.
Therefore, and unlike the nearest neighbor FM Heisenberg model, the
spin-wave dispersion does not depend linearly on the coupling amplitudes
$\{J_x,J_z,I\}$, but as in the AF Heisenberg case has a square-root form:
\bea
\label{disFx}
\omega_{Fz}(\vec k)\! &=&\! 4S \sqrt{\big(2|J_z+2I|+I_{\vec k}
+  J_{kx}\big)^2 - J_{kx}^2  },\;\;\;\; \\
\{J_{kx}, I_{\vec k}\}\! &=&\! \{J_x \cos k_x, 2I(\cos k_x+\cos k_z )\}.
\label{eq:swfz}
\eea

\begin{figure}[t!]
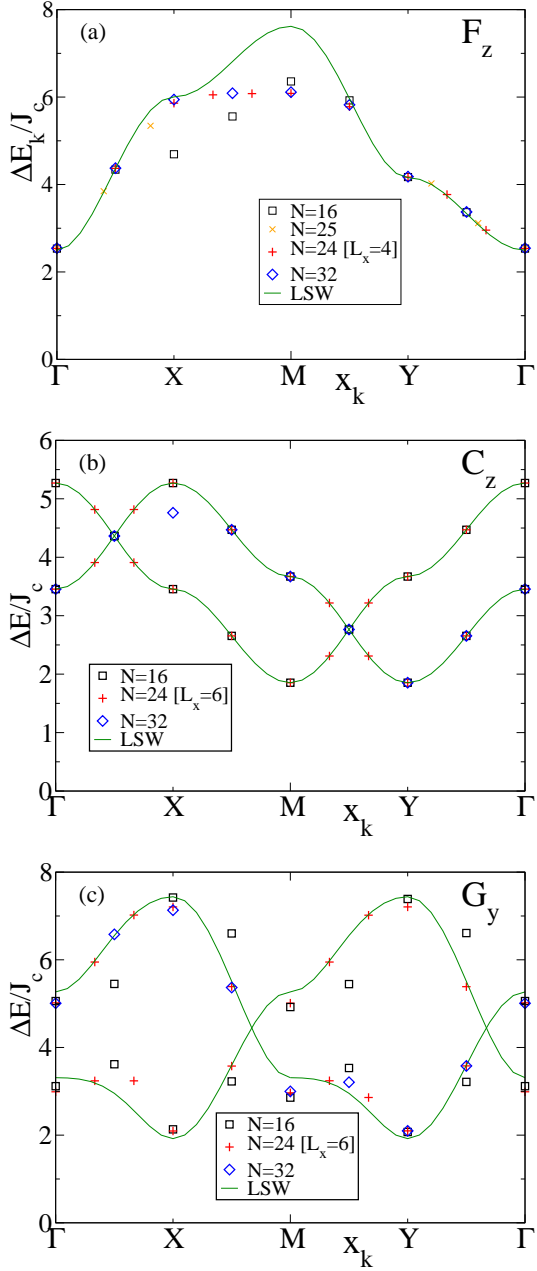

\begin{center}
\includegraphics[width=7cm]{swaves_Im02t23.eps}\\
\vspace{0.3cm}
\includegraphics[width=7cm]{swaves_I02t23.eps}\\
\vspace{0.3cm}
\includegraphics[width=7cm]{swaves_I06t23.eps}\\
\caption{\label{fig:sw1} (Color online)
Spin wave dispersions obtained within the LSW theory (lines) and by exact
diagonalization (symbols) of  various clusters with size $N \leq 32$
for $\phi=23\pi/20$ and for different Heisenberg coupling constants:
(a) $I=-0.2J_c$; (b) $I=0.2J_c$; and (c) $I=0.6J_c$. These different
parameters correspond to the $F_z$, $C_z$ and $G_y$ phases,
respectively. The abscisse $x_k$ follows a path in the Brillouin zone
identical as in Fig.~\ref{fig:sq2}.
For $N=32$ only one of the two lowest spin-wave branches
[see second paragraph below Eq.~(\ref{eq:swgz})] is shown.
}
\end{center}
\end{figure}

The dispersion Eq.~(\ref{disFx}) is shown in Fig.~\ref{fig:sw1}(a) on a
closed path $\Gamma \rightarrow X \rightarrow M\rightarrow Y\rightarrow
\Gamma$ in the first Brillouin zone for the coupling constants:
$I=-0.2J_c$ and $\phi=\pi+3\pi/20$. The LSW approximation describes well
the lowest excitation energies obtained by exact diagonalization
(shown in the same figure for several periodic clusters).
In diagonalization spin waves can be identified by the parity of
$\sum_{\vec r} S^z_{\vec r}$ that is opposite to that of the GS. One
notices that for momenta such that $\omega({\vec k})$ exceeds a certain
critical value, spin-wave energies obtained by exact diagonalization
tend to differ from the LSW results; and this critical value seems to increase 
with system size. These features are actually related to the presence, in the 
low-energy spectrum of finite clusters, of the column-flip excitations which
will be analyzed in Section \ref{sec:cex}. When the energy $\propto L_z|I|$ of
such excitations coincides with that of spin
waves, their interaction leads to deviations from the LSW dispersion.
Note that in general the spin waves obtained in the LSW theory are
gapped, except in two limits: $J_x/I, J_z/I \rightarrow 0$
(one recovers here the dispersion of the nearest neighbor Heisenberg
ferromagnet),
and for $J_x=J_z$. The latter softening, occurring at $\vec k = \Gamma$,
is associated to the transition to the $F_x$ phase.

We turn now to the description of spin waves in AF phases. Here, the
vacuum of Holstein-Primakoff bosons has to be defined differently; one
can apply a transformation similar to those seen in Sec.~\ref{sec:sym},
that is, inverting two components of spins on one sublattice, in order
to obtain (after this transformation) a Hamiltonian with FM classical
ground states. A concrete example is given here with the $C_z$ phase
found in the phase diagram of Fig.~\ref{fig:phdF}. Here such a
transformation consists of inverting $y$ and $z$ spin components only on
columns with $j$ even. After this, not only compass- but also Heisenberg
couplings on $x$ bonds contribute to $a_{\vec r}\,a_{\vec s}$-type terms
in the LSW Hamiltonian. The resulting dispersion is:
\beq
\omega_{Cz}(\vec k) = 4S \sqrt{\big(2|J_z| + J_{kx} + I_{kz}\big)^2
  - \big(J_{kx}+I_{kx} \big)^2}.
\label{eq:swcz}
\eeq
In the previous expression and hereafter,
\beq
\{J_{k\alpha},I_{k\alpha}\}=\{J_\alpha \cos k_\alpha, 2I\cos k_\alpha\}.
\label{eq:parc}
\eeq
Similarly, we derive the dispersion found by LSW approximation for the
lowest spin waves in N\'eel-like phases of the model, namely the $G_y$
and $G_z$ phase:
\bea
\label{eq:swgy}
\omega_{Gy}(\vec k) &=& 4S
\sqrt{(I_{\vec 0}+J_{kz}-J_{kx})^2-(I_{\vec k}+J_{kx}+J_{kz})^2}, \nonumber \\
\\
\omega_{Gz}(\vec k) &=& 4S \sqrt{(2J_z+4I+J_{kx})^2-(J_{kx}+I_{\vec k})^2}.
\label{eq:swgz}
\eea
Here we use parameters defined Eq.~(\ref{eq:swfz}) and Eq.~(\ref{eq:parc}).
To illustrate these dispersions in the $C_z$ and $G_y$ phases, we show
them along the path
$\Gamma \rightarrow X \rightarrow M \rightarrow Y \rightarrow \Gamma$
in Fig.~\ref{fig:sw1}(b) (for $I/J_c= 0.2$) and Fig. \ref{fig:sw1}(c)
(for $I/J_c= 0.6$), respectively, with the same anisotropy parameter
$\phi=23/20$ in both cases. In Fig.~\ref{fig:sw1}(b) the correspondence
between numerical results and the dispersion Eq.~(\ref{eq:swcz}) is
remarkable. Only in the energy range $\Delta E_k\gtrsim 5J_c$ one
notices a tiny discrepancy between the numerical and LSW results. This
can be due to the interaction of single-magnon excitations with
column-flip excitations at energy $\propto L_zI$, as in the $F_z$ phase.
In Fig.~\ref{fig:sw1}(c) the agreement between the LSW expression for
the $G_y$ phase and the numerical results is also satisfactory,
although finite-size effects are larger than in the case of
Fig.~\ref{fig:sw1}(b). The deviations from LSW theory for the $G_y$
phase are not due to column-flip excitations (these are
not properly defined in the $G_y$ phase) but are rather induced by
the proximity of the $C_z\leftrightarrow G_y$ transition.

In the phases corresponding to Figs.~\ref{fig:sw1}(b) and
\ref{fig:sw1}(c) and more generally in AF spin-gapped phases of the
CH model, the lowest LSW branch is actually doubled due to the twofold
ground state degeneracy in the classical limit:
each branch contains the lowest spin waves above a linear combination
of classical ground states, which belongs to a given symmetry
representation. In the example of the $C_z$ phase, this representation
can be of momentum either $\vec k=\Gamma$ or $\vec k = X$, and in the
latter case the associated LSW dispersion is $\omega_{Cz}(\vec k + X)$
instead of $\omega_{Cz}(\vec k)$. A similar branch doubling occurs in
N\'eel-like phases, resulting for the $G_\alpha$ phase in a second
branch of dispersion $\omega_{G\alpha}(\vec k + M)$ along with that of
dispersion $\omega_{G\alpha}(\vec k)$.
In FM phases, branches are also doubled as a consequence of the
$\mathbb{Z}_2$ symmetry of the ground state. This doubling does not appear on
dispersion plots like the ones shown in Fig.~\ref{fig:sw1}(a) since the two
LSW branches, above ground states of identical momentum $\Gamma$, have the
same dispersion. Nevertheless, and similarly as in AF phases, each
spin-wave excitation found in exact diagonalization can be attributed to
one of these branches, thanks to its parity even or odd under
time-reversal symmetry (or equivalently, in a phase with easy axis
$\alpha$, under the symmetry
$\sigma^\alpha_{\vec r} \rightarrow-\sigma^\alpha_{\vec r}$).
We plot both LSW branches in these figures; note also that one can
deduce which translational symmetry is broken in the TL from the
relative position in momentum space of both branches [e.g. the shift of
$\vec q=X$ between both branches evidences a breaking of translation
symmetry by $\vec{e}_x$ in Fig.~\ref{fig:sw1}(b)].

\begin{figure}[t!]
\begin{center}
\includegraphics[width=7.7cm]{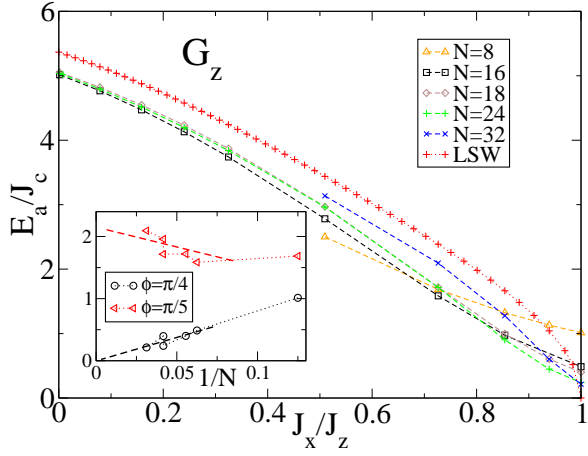}
\caption{\label{fig:sw2} (Color online)
Anisotropy gap $E_a/J_c$ of spin waves in the $G_z$ phase as function of
increasing $J_x/J_z=\tan(\phi)$, for Heisenberg amplitude $I=J_c/5$ and
various cluster sizes $N$. Inset: Size-scaling for the gap for:
$\phi=\pi/4$ (isotropic case) and for $\phi=\pi/5$ (finite gap).}
\end{center}
\end{figure}

In an ordered phase of the model,
the minimum of the dispersion of the lowest spin wave branch ---
or branches if taking into account the branch doubling --- is
important for two reasons. First, the corresponding excitation energy
(or spin gap) is to be compared to the energy $\propto L_zI$ of
excitations mentioned above, which require a more detailed description
given in the next Section. Second, when varying parameters of the model
this spin gap vanishes at transitions with other phases, provided the
transition does not occur on the compass line $I=0$. A good example is
the case of the $G_x\leftrightarrow G_z$ transition, when $J_x$ is varied
while $I$ and $J_z$ are kept fixed and finite.
In the $G_z$ phase the gap to spin waves $E_a$ is, for finite clusters,
the lowest excitation energy in the sector of odd $\sum_{\vec r} S^z_{\vec r}$,
found in representations of either $\vec k=\Gamma$ or $\vec k = M$. It
is shown in Fig.~\ref{fig:sw2} along with the corresponding LSW prediction
from Eq.~(\ref{eq:swgz}), in function of $J_x/J_z$. Even though finite-size
effects are not negligible away from the transition, i.e., for $J_x<J_z$,
attempted scalings clearly indicate finite spin gap values in the TL which
is comparable to the LSW theory result. Instead,
at $J_x=J_z$, such a scaling confirms that there the spin gap vanishes
in the TL. Within the LSW theory one finds that the spin gap vanishes at
the transition as $c\sqrt{|J_z-J_x|}$. The symmetry relation connecting
each point of the $G_z$ phase, in the phase diagram Fig.~\ref{fig:phdAF},
to a point of the $G_x$ phase and vice-versa, implies a relation between
spin-wave dispersions in both phases; the LSW result for the $G_x$ phase
is given by inverting, in Eq.~(\ref{eq:swgz}), first $J_x$ and $J_z$,
and second $k_x$ and $k_z$ (these dispersions being even functions of
$k_x$ and $k_z$).

One finds similar mode softening at the transition between $C_z$ and
$G_y$ phases, although these phases are not related to each other by
any exact mapping. Here, if one approaches the transition from the side
of the $C_z$ phase, by increasing $|I/J_z|$ with fixed $J_x/J_z\in (0:1)$,
the dispersion $\omega_{Cz}(\vec k)$ has a minimum at $\vec k=Y$
according to Eq.~(\ref{eq:swcz}), see Fig.~\ref{fig:sw1}(b),
which vanishes for $I/J_z=-1/2$. Similarly, in the $G_y$ phase and close
enough to the transition towards the $C_z$ phase, the dispersion given by
Eq.~(\ref{eq:swgy}) is characterized by a minimum at $\vec k=X$ (the other
branch $\omega_{Cz}(\vec k+M)$ has a minimum of equal value at $\vec k=Y$)
which vanishes at $I=-J_z/2$ as well. Note that the closeness to one
another of spin gap values $E_a=\min_{\vec k}\omega(\vec k)$ in the
three cases shown in Fig.~\ref{fig:sw1} is accidental and is merely a
consequence of the choice of the $I/J_z$ value for each case. The
cases of $C'_z\leftrightarrow F_y$ and $C'_z\leftrightarrow F_x$ transitions,
addressed in the previous Section, are characterized by similar mode
softening, qualitatively reproduced within the LSW theory;\cite{clasw}
in each phase the spin gap corresponds (up to finite-size effects) to the
lowest excitation energy seen in Figs.~\ref{fig:eextr} and
\ref{fig:eextr2} at $\vec k=Y$ ($F_y$ or $F_x$ phases), or at
$\vec k=\Gamma$ ($C'_z$ phase). Interestingly, in an ordered phase
$\Phi$ but sufficiently close in the phase diagram to a transition
line (except the line at $I=0$) to another phase, the momentum
$\vec k_0$ corresponding to the minimum in the LSW dispersion
$\omega_\Phi(\vec k)$ allows one to deduce which translation
symmetries are spontaneously broken in this other phase.

\section{Column-flip excitations on nanoclusters}
\label{sec:cex}

In the CH model there is another distinct set of elementary excitations,
i.e., in addition to the spin-waves. These are the column-flip
excitations that correspond to a reversal of all spins in a column of
strongly coupled spins in the case $|J_z|>|J_x|$ and small Heisenberg
amplitude $|I|$.\cite{rows} These excitations emerge from the
macroscopic ground state degeneracy of the original compass model
and reflect the twofold ground state degeneracy of ordered phases
selected by the Heisenberg interactions.
Whereas spin-wave excitations yield essentially the same energy of order
$O(1)$ for small clusters (within exact diagonalization) and in the TL,
this is distinct for the column-flip excitations whose energy scales
with a linear dimension of the system. In the following paragraph we
will analyze the \textit{column-flip excitations} by means of an
effective pseudospin model which we will derive from the original CH
model using high-order perturbation theory. Then we will employ this
effective model and will show how a quantum computation scheme involving
the column-flip excitations can be conceived. This requires the
fulfilment of certain conditions on the low-energy excitation spectrum,
which we will eventually examine.

\subsection{Derivation of an effective columnar model}
\label{sec:eff}

\begin{figure}[t!]
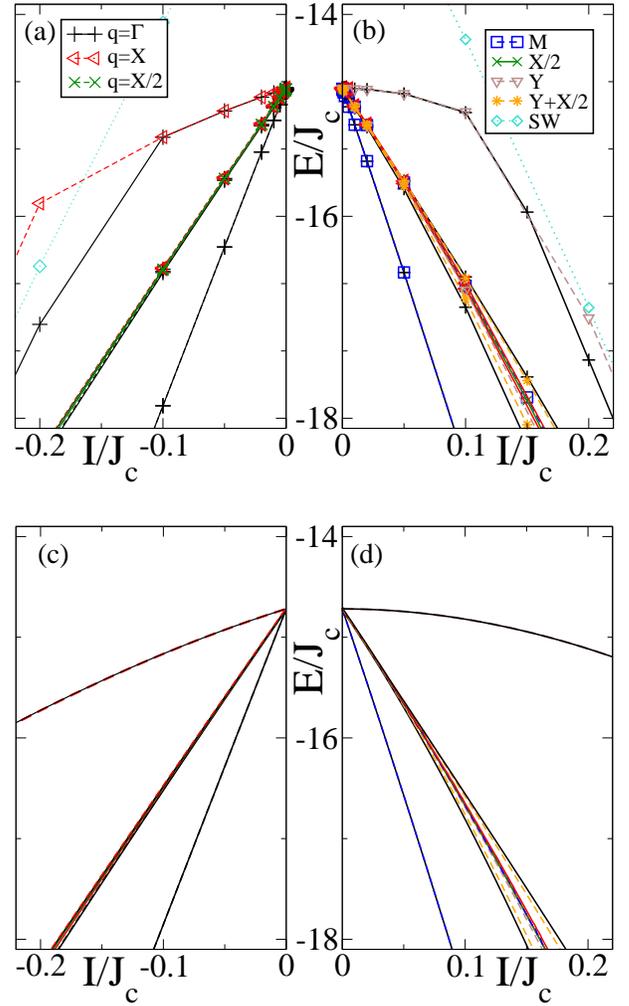

\begin{center}
\includegraphics[width=8cm]{spectrext3.eps}\\
\vspace{0.5cm}
\includegraphics[width=8cm]{spectrefft3.eps}\\
\caption{\label{fig:exef} (Color online)
Low-energy spectra of columnar excitations $E/J_c$ obtained by exact
diagonalization of a $L_x=L_y=4$ periodic cluster, as function of
$I/J_c$, for:
(a) FM interactions ($\phi=23\pi/20$), and
(b) AF interactions ($\phi=3\pi/20$), respectively.
Corresponding results obtained from the effective Hamiltonian
$H_{\rm col}^{\rm FM/AF} + H_{\rm col}^{\rm FM/AF'}$ for the same
cluster and interaction parameters are shown for the FM case (c) and AF case
(d), respectively. Momenta of various eigenstates are indicated by symbols;
continuous and dashed lines indicate even and odd states w.r.t. time
reversal.
The lowest line in each panel represents the
two quasi-degenerate ground states,
while the next band ({\it central branch}) contains  $2^4-4=12$ columnar
excitations.
}
\end{center}
\end{figure}

We consider here finite clusters of size $L_x\times L_z$ with
anisotropic CH interactions, assuming $|J_z|>|J_x|$ without loss of
generality. The amplitude $I$ of Heisenberg interactions is chosen
finite but small compared to $|J_z|$, as this
is known (see Sec.~\ref{sec:s(q)}) to be sufficient to lift the
quasi-degeneracy between
$2^{L_x}$ \textit{columnar states}. This splitting implies that, among
those states, the $(2^{L_x}-2)$ ones which do not correspond to the
ground states of the selected ordered phase acquire finite excitation
energies due to the finite value of $I$. Figure \ref{fig:exef}(a) shows
energies of lowest eigenstates as function of $I$ for a
square cluster with edge length $L_x=L_y=4$. For $I$ small enough
($\lesssim 0.15$) a group of 16 distinct eigenstates lies below the
lowest spin-wave excitation --- these 16 states have energies varying
roughly linearly with $I$, with different branches corresponding to
different slopes $dE/dI$. In the AF case ($J_z>0$) one sees in Fig.
\ref{fig:exef}(b) the same type of excitation branches, with
energies depending linearly on $I$ when this quantity is small
enough. The main difference is the somewhat larger splitting of
excitation energies within a multiplet-branch in the AF case.

We have seen in Sec.~\ref{sec:chm} that an effective Hamiltonian
$H_{\rm col}^{(0)}$ provides a valuable insight into the QCM ($I=0$)
for the anisotropic case $|J_z|>|J_x|$. This effective model uses a
formalism of pseudospins $\{\vec{\tau}_j\}$ describing the columnar
states forming the low-energy subspace.
Here, to describe the peculiar properties of the low-energy spectrum in the
case where both $I$ and $J_x$ are finite but small w.r.t. $|J_z|$,
we will derive a more general effective Hamiltonian $H_{\rm col}$,
expressed in the same pseudospin formalism. In the derivation we consider
compass couplings on $x$ bonds and Heisenberg couplings as perturbations.
In the FM case ($J_z<0$), the resulting effective model is a 1D XYZ
Hamiltonian in terms of pseudospins $\vec{\tau}_j$,
\bea
\label{efffm}
H^{\rm FM}_{\rm col}=-N|J_z| + \sum_{j=1\ldots L_x} \sum_{\alpha \in {x,y,z}}
C^\alpha \tau_j^\alpha\tau_{j+1}^\alpha ,
\eea
where the coupling constants are given by:
\begin{eqnarray}
C^z&=& L_z I , \\
C^{x/y}&=& -\frac{J_{\rm col}}{2}
\left\{\left(1+\frac{2I}{J_x}\right)^{L_z} \pm 1\right\}.
\label{eq:efJxI}
\end{eqnarray}
Here $J_{\rm col}$ is given by Eq.~(\ref{eq:efJx}) and depends again on
cluster size and boundary conditions. The constant $-N|J_z|$ is the
ground state energy of the unperturbed Hamiltonian. This effective
\textit{columnar Hamiltonian} describes the structural and (quantum)
dynamic properties of the low-energy, column-ordered states in CH
nanoclusters.

One can qualitatively interpret the difference between
the Hamiltonian Eq. (\ref{efffm})
and $H^{(0)}_{\rm col}$ obtained previously for $I=0$, by listing the
various roles played by Heisenberg couplings in the perturbation theory:\\
(i) Most important are the $I\sigma^z \sigma^z$
couplings on horizontal bonds --- they split the degeneracy of columnar
states at first order in perturbation theory and contribute to the terms
$\propto L_z I\tau_j^z\tau_{j+1}^{z}$ which account for the ordering in
the TL discussed in Sec.~\ref{sec:s(q)}.\\
(ii) The couplings $I\sigma^z \sigma^z$ on vertical bonds,
instead, do not distinguish between the columnar states, but they
contribute to their energy, either by a quantity $L_x L_z I$ (with PBC)
or $L_x(L_z-1)I$ (with OBC).\\
(iii) The transverse components $2I(\sigma^+\sigma^- +{\rm H.c.})$
of Heisenberg couplings on horizontal bonds have to be added to terms
$J_x\sigma^x\sigma^x$ when evaluating the transverse coupling amplitudes
$C^x$ and $C^y$. Here, not only $\tau^x\tau^x$ terms but also (smaller)
$\tau^y\tau^y$ terms appear at order $L_z$ in perturbation theory.\\
(iv) Eventually, the transverse Heisenberg couplings
$2I(\sigma^+\sigma^- +{\rm H.c.})$ on vertical bonds have to be
considered  {\it a priori} in the perturbative approach. For $J_z<0$
they can be left out, since columnar states have spins ferromagnetically
aligned within columns; but for $J_z>0$ these couplings
allow for effective single-column flips, which appear at order $L_z/2$ in
perturbation.
As a result the effective Hamiltonian $H^{\rm AF}_{\rm col}$ is formally
the sum of $H^{\rm FM}_{\rm col}$ and of an additive term:
\bea
H^x_{\rm col}&=& -I_{\rm col} \sum_j \tau_j^x,  \\
I_{\rm col} &=& L_z\,2^{L_z/2}\gamma^{(')}_{L_z/2}
J_z\,\left(\frac{I}{2J_z}\right)^{L_z/2},
\label{eq:icol}
\eea
accounting for a single column flip. It appears at order
$L_z/2$ in perturbation.

The strength $|C^y|$ of the $\tau^y\tau^y$ coupling is, for $0<|I|\ll
|J_x|,|J_z|$, much smaller than that $|C^x|$ of the $\tau^x \tau^x$ coupling;
the former vanishes for $I \rightarrow 0$, where one recovers the effective
Hamiltonian $H_{\rm col}^{(0)}$. Both coupling strengths are, for $L_z$ large
enough, much smaller than the one $|C^z|$ of the $\tau^z \tau^z$
coupling. This allows us to split columnar states for the latter coupling, so
that the low-energy spectrum has the branch-like structure seen in
Fig.~\ref{fig:exef}(c). For this system size there are three branches
corresponding to the twofold degenerate ground state, the central one and the
upper multiplet branch. These energy splittings are given by the number of
domain walls, i.e., 0, 2, or 4 in the pseudospin chain.

We comment here briefly to the implications of this effective model for
the interpretation of finite size data in previous Sections, in
situations where $|I|\ll |J_x|<|J_z|$. There, we stated that the reason
why ordered phases are favored even with infinitesimal Heisenberg
couplings appears clearly with this effective description. And indeed,
from Eqs.~(\ref{eq:efJxI}) and (\ref{eq:icol}) it is clear that while
parameters $C^{x/y}$ and $I_{\rm col}$ vanish exponentially in the TL,
$C^z$ does not (and even diverges in this limit). More quantitatively,
for finite clusters one can expect that the order favored by effective
$\tau^z\tau^z$ couplings occurs when $|C^z|>|C^x|$ and
$|C^z|>|I_{\rm col}|$. Here the example shown in Fig.~\ref{fig:sq2},
for $I/J_c=10^{-6}$, $\phi=\pi/10$, and $L_x=L_z=6$, is instructive:
the corresponding effective coupling amplitudes are given by
$I_{\rm col}/(L_zJ_z)\simeq 1.0\times 10^{-18}$,
$J_{\rm col}/(L_zJ_z)\simeq 1.4\times 10^{-6}$, and
$C^z/(L_zJ_z)\simeq 1.0\times 10^{-6}$.
Thus $|C^x|$ is somewhat larger than $C^z$, explaining the broad
distribution of $S^z(\vec k)$ over all momenta such that $k_z=\pi$,
as on the compass line; but even for such small Heisenberg amplitude
$I$, $C^z$ is not negligible relative to $|C^x|$ and in
consequence $S^z(\vec k)$ is much larger at $\vec k = M$ than at other
wave vectors. Similarly, the sharp peaks at $I=0$ in the fidelity plots
of Fig.~\ref{fig:fid} indicate that the Heisenberg coupling strength
necessary for the $C'_z$ or $G_z$ order (depending on the sign of $I$)
to develop on clusters considered is even smaller than the resolution
$0.005J_c$ chosen in that plot. According to the criterion $C^z\gg C^x$
(with $I_{\rm col}$ negligible in this regime) for $\phi=-3\pi/20$ even
a value $|I|\gtrsim 10^{-3}J_c$ is sufficient to develop long-range order.

Until now we only considered the effective couplings found in leading
order in perturbation for each pseudospin component $x,y,z$, and
explaining the overall structure of the low-energy spectra in
Figs.~\ref{fig:exef}(a) and \ref{fig:exef}(b). For a more accurate
description
of those, we include in the effective Hamiltonian, besides
$H^{\rm FM/AF}_{\rm col}$, subleading terms $H^{\rm FM/AF'}_{\rm col}$
found at second order in perturbation theory.
The latter account for processes where two spins are flipped on nearest
neighbor bonds. For $J_z<0$ only horizontal bonds have to be considered;
since the amplitude of these terms, for $I\ne 0$, depends on whether the
pseudospins involved are parallel (amplitude $J_x$) or antiparallel
(amplitude $J_x+2I$), the resulting nearest neighbor coupling is of the
$\tau^z\tau^z$ type. The part of the effective Hamiltonian stemming from
these processes is (here for $J_z<0$):
\beq
H^{\rm FM'}_{\rm col}= -N\frac{J_x^2}{8|J_z|} - \frac{I(J_x+I)}{4|J_z|}
\sum_j L_z(1-\tau^z_j \tau^z_{j+1}).
\label{eq:h2eff}
\eeq
The spectra in Figs.~\ref{fig:exef}(c) and \ref{fig:exef}(d) correspond
to effective Hamiltonians $H^{\rm FM}_{\rm col}+H^{\rm FM'}_{\rm col}$
and $H^{\rm AF}_{\rm col}+H^{\rm AF'}_{\rm col}$, respectively --- in
the latter case $H^{\rm AF'}_{\rm col}$ differs from
$H^{\rm FM'_{\rm col}}$ by the presence of an additional constant
$-NI^2/J_z$, accounting for fluctuations on vertical bonds.
The inclusion of $H^{\rm FM/AF'}_{\rm col}$ leads to a much better
agreement with the original spectra of Figs.~\ref{fig:exef}(a) and
\ref{fig:exef}(b), regarding absolute energies and their dependence on
$I$, than if diagonalizing $H^{\rm FM/AF}_{\rm col}$ alone. For instance,
adding this correction term one reproduces that the dependence of the
energies of central and upper branches on $I$
is not simply linear but contains a (small) quadratic contribution.

This effective model gives not only a good estimate of lowest excitation
energies, but also the correct quantum numbers for the corresponding
eigenstates within each branch. The two states of the lowest branch in
Figs.~\ref{fig:exef}(a) and \ref{fig:exef}(c), which become the twofold
degenerate ground states of the $F_z$ phase in the TL, obviously both
have momentum $\Gamma$. Similarly, the two eigenstates forming the
highest branch are linear combinations of states with a $C_z$-like
pattern, one at $\vec q = \Gamma$ and the other at $\vec q = X$.
The remaining, intermediate branch in Fig.~\ref{fig:exef}(a) corresponds
to the subspace generated by $2^4-4=12$ columnar states such that, in
terms of pseudospins, $\sum_i \tau^z_i \tau^z_{i+1}=0$: this branch contains
$|\uparrow_1\uparrow_2\downarrow_3\downarrow_4\rangle$,
$|\uparrow_1\uparrow_2\uparrow_3\downarrow_4\rangle$, and
$|\uparrow_1\downarrow_2\downarrow_3\downarrow_4\rangle$,
plus for each of those the three states obtained by translations along
the pseudospin chain. The effective model allows us to understand why three
eigenstates of this branch are found in each representation of momentum
such that $q_z=0$. Also the dispersion within a branch is well
reproduced by the effective model, however the splittings are too small
for this dispersion to be visible in the spectra of Fig.~\ref{fig:exef}.

In the AF CH case (for $J_z>0$), the presence in the effective
Hamiltonian of $H_{\rm col}^x$ has a significant impact on the properties of
the intermediate branch, clearly visible in Figs.~\ref{fig:exef}(b) and
\ref{fig:exef}(d): since the amplitude $I_{\rm col}$ is much larger than
$C^{x/y}$ for $I$ large enough ($I \gtrsim 0.05J_c$ in the present
example), the energy dispersion within this branch is dominated by
$I_{\rm  col}$ in this interaction range, and much broader than
in the FM case [Figs.~\ref{fig:exef}(a) and \ref{fig:exef}(c)] for equal
value of $|I|$.

We make here two important remarks on the generalization of features
above to other clusters:\\
(i) In the previously discussed case of a periodic cluster with
$L_x=4$, only one intermediate branch with a large quasi-degeneracy is
found. This is specific to this case, and for clusters with larger $L_x$
one has several such branches --- for instance with $L_x=6$ and PBC one
has two of them; each one is generated by 30 columnar states where
$\sum_i\tau^z_i\tau^z_{i+1}$ takes the value $-2$ or $+2$ respectively.
The maximal number of states per branch, which is given by $2C_{L_x}^p$
with $p$ (one of) the even integer(s) closest to $L_x/2$, grows
exponentially with the number of columns.\\
(ii) The obtained branch structure of the low-energy spectrum is a
property of finite and small enough clusters, but does not require
periodic boundaries. We have verified that
for open clusters, as for periodic ones, the description of these systems
with perturbative techniques is possible and
efficient, but the obtained effective amplitudes for transverse couplings
$\tau_j^\alpha \tau_{j+1}^\alpha$ are modified, being proportional to
$\gamma'_{L_z}$ instead of $\gamma_{L_z}$. The amplitudes 
of dominant $\tau^z \tau^z$ couplings are the same as for PBCs, but the
number of branches that split off is larger here ($L_x-1$) since odd
numbers of domain walls are allowed, in contrast to the periodic case.

\subsection{Quantum computing scheme based on quasi-degenerate columnar states}
\label{sec:qccol}

We have seen that the low-energy spectrum of open $L_x\times L_z$ clusters
where $\sigma$-spins interact via CH couplings can be
well reproduced, for small enough Heisenberg amplitudes and sufficiently
anisotropic compass couplings, by an effective XYZ-type model of
pseudospins $\tau=1/2$. A remarkable feature of this
spectrum is the subdivision of the $2^{L_x}$ columnar states, forming a
low-energy subspace selected by dominant compass interactions, in
multiplet branches of quasi-degenerate states. Some branches have a
semi-macroscopic degeneracy, while the lowest branch contains only the
two degenerate ground states selected by Heisenberg perturbations.
The ground states are characteristic for the order of the respective phase
and the $\mathbb{Z}_2$-symmetry of the model. The
effective model allows, thanks to the high
anisotropy of its coefficients ($|C^z|\gg |C^x|,|C^y|$), for a good
understanding of the number and the nature of
columnar excitations in the different branches.

\begin{figure}[t!]
\begin{center}
\includegraphics[width=7.5cm]{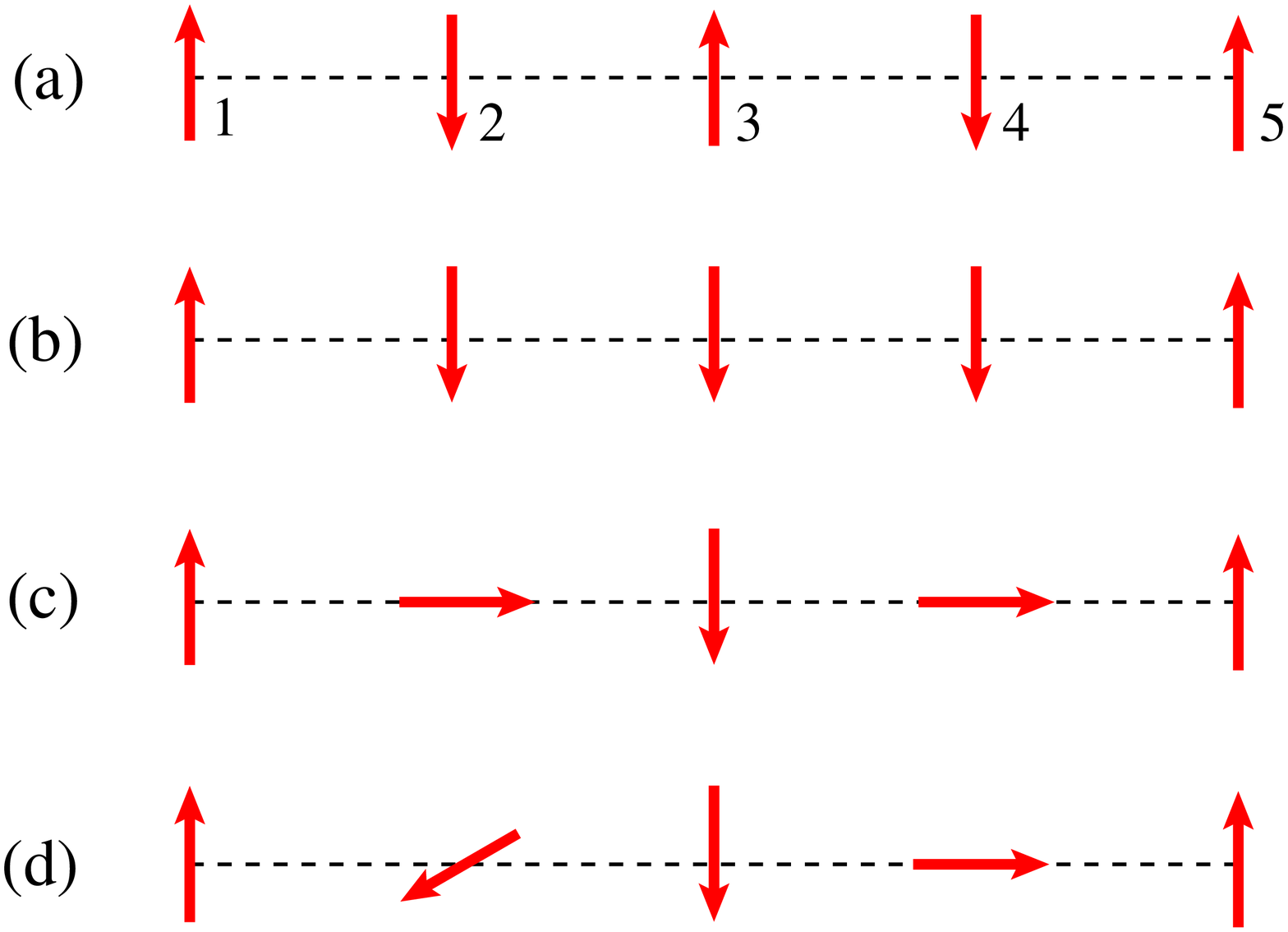}\\
\caption{\label{opecol}
Pseudospins of a CH nanocluster representing
$L_x=5$ columnar spin chains in the case $|J_z|\gg|J_x|, |I|$.
Panels (a)-(d) show the successive steps in the initialization and
utilization of two essentially decoupled protected qubits:
(a) the system is in one of the two quasi-degenerate ground states,
$|\Phi_0\rangle$,
such that for each pseudospin $j$ one has  $\tau_j^z=(-1)^{j-1}$;
(b)~after flipping the column (or pseudospin) $j=3$, the system is in
an excited state $|\Phi^*\rangle$ belonging to the \textit{central}
branch of columnar excitations;
(c) the qubits encoded by pseudospins $\vec{\tau}_2$ and $\vec{\tau}_4$
are initialized in the state $\rm{H}_2 \rm{H}_4|\Phi^*\rangle$, where
$\rm{H}_j$
rotates the pseudospin $\vec{\tau}_j$ along the $y$ axis in pseudospin
space;
(d) example of a state produced from the former by an operation on
pseudospin $\vec{\tau}_2$.}
\end{center}
\end{figure}

This suggests that one can control the quantum state of the system and
possibly realize elementary operations of quantum computing by using a
subspace of quasi-degenerate states. The starting point is to excite the
system into a given branch, by flipping given columns; then one can
initialize the quantum computer by placing the system in a state where
some pseudospins (the qubits of the computer) are highly entangled; and,
after this, perform quantum operations by acting on these qubits.
For concreteness, we specify a system of interest: a rectangular, open
cluster of $L_x\times L_z$ spins (we set in our example $L_x=5$ for the
rest of this paragraph). Concerning interaction parameters, we choose FM
compass ($J_x<J_z<0$) and AF Heisenberg ($I>0$) couplings,
which corresponds to the columnar $C_z$ phase in Fig.~\ref{fig:phdF}.
This choice
is, within the framework of the CH model, the closest one to possible
realizations using arrays of (pseudo)spins coupled by dipolar
interactions, see Eq.~(\ref{eq:Hdip}). One of the two lowest
eigenstates of the system (these are quasi-degenerate ground states) can
be expressed, as in Fig.~\ref{opecol}(a), in terms of pseudospins
$\vec{\tau}_i$ by:
\beq
|\Phi_0\rangle=|\uparrow_1 \downarrow_2
\uparrow_3 \downarrow_4 \uparrow_5\rangle.
\eeq
By flipping the $j=3$ column, i.e., the pseudospin $\vec{\tau}_3$,
one obtains the state shown in Fig.~\ref{opecol}(b):
\beq
|\Phi^*\rangle=|\uparrow_1 \downarrow_2
\downarrow_3 \downarrow_4 \uparrow_5\rangle.
\eeq
This state belongs, in a limit where $C^{x/y} \ll C^z$,
to a branch of 12 quasi-degenerate states; a basis of this
branch consists of eigenstates of $\tau^z_j$ operators, such that
$\tau^z_1=\tau^z_5=\pm 1$ and
for exactly two values $j \in [1;4]$ one has $\tau^z_j \tau^z_{j+1}=-1$.
We will consider from now on operations within this subspace, which we
call \textit{central branch} since the number of domain walls is half of
the maximum allowed $(L_x-1)$.

Such a manifold of quasi-degenerate states can a priori be used for the
implementation of a quantum computing scheme; but a necessary condition
is that the intrinsic quantum dynamics should not interfere with
the computation process. To this end, we further restrict the
\textit{work subspace}, i.e. the subspace spanned by all possible states
available via operations on qubits, by imposing that the pseudospins
$\vec{\tau}_1$, $\vec{\tau}_3$ and $\vec{\tau}_5$ remain in the same
state as in $|\Phi^*\rangle$. This leaves 2 pseudospins $\vec{\tau}_2$
and $\vec{\tau}_4$ which can be operated by pulses acting coherently on
all spins of the corresponding column, while keeping the system in this
branch of states. One can first initialize this 2-qubit computer by
applying Hadamard gates on pseudospins $\vec{\tau}_2$ and $\vec{\tau}_4$,
resulting in the following state [shown on Fig.~\ref{opecol}(c)]:
\beq
\rm{H}_2 \rm{H}_4|\Phi^*\rangle=|0\rangle=\frac12
\sum_{\tau_2^z,\tau_4^z \in {\uparrow;\downarrow}} |\uparrow_1 \tau_2^z
\downarrow_3 \tau_4^z \uparrow_5 \rangle,
\eeq
with $\rm{H}_j$ corresponding to the operator
$(\tau_j^x+\tau_j^z)/\sqrt{2}$. Further actions on either of the columns
$j=2$ and $j=4$ (such that the system remains in the work subspace) will
correspond to elementary operations of the corresponding qubits, leading
to intermediate states as the one represented on Fig.~\ref{opecol}(d).
This scheme can be extended to systems with more qubits, namely $p$ when
using an open cluster with $L_x=2p+1$ columns; in this more general case
$|\Phi^*\rangle$ is an eigenstate of all $\tau^z_j$ operators such that
$\tau^z_j\tau^z_{j+2}=-1$ for all
columns of index $j=2k-1$ ($k \in [1;p]$); qubits are then encoded in the
remaining columns with $j$ even.

This encoding scheme presents several advantages: first the qubits are
semi-locally defined, i.e., each qubit corresponds to degrees of freedom
of a specific column and can be manipulated by a field acting on this
column. Second, the structure of columnar excited states makes them
robust against local noise. When considering local perturbations to the
CH Hamiltonian, for instance of the form
$\sum_{\vec r} h_{\vec r} \sigma_{\vec r}^z$ with variables $h_{\vec r}$
randomly distributed (and small),
\cite{Dou05} these can reverse columns only at order $L_z$ in
perturbation theory. Qubits based on columnar excitations are thus
intrinsically, as in the case of topological quantum computing, much more
robust to such perturbations than qubits which would be defined locally,
e.g. on a single spin. Besides, the choice of using the work subspace
described above has a further advantage. To see this, we consider the
operator $Q_j$ that flips the whole column $j=2k$ where the $k$th qubit
is encoded; the commutator of this operator with $\cal{H}$ is given by
Eq.~(\ref{comut}). From this we see that within the work subspace the
$\alpha=z$ term of the sum vanishes
(since in this subspace
$\sigma_{i,j-1}^z + \sigma_{i,j+1}^z=0$ on all rows $i$).
Thus expectation values of $[Q_j,\cal{H}]$ are expected to
be smaller within this subspace than within any other quasi-degenerate
branch of columnar states. Therefore the scheme described here appears as
the optimal way to encode qubits using low-energy eigenstates of CH
nanoclusters.

\subsection{Protection against relaxation and decoherence}

The quantum computation scheme presented above would work perfectly if:
(i) within the branch involved in the scheme, eigenstates were
\textit{exactly} degenerate, and
(ii) the quantum gates realizing the transition from one state to
another could be implemented perfectly by physical operations.
The latter condition means that one could apply a spatially-resolved
magnetic field, focused on a given column of spins, for a fixed time
$t_1$; and, noting the initial state $|A\rangle$ (belonging to the work
subspace) at $t=0$, the state $e^{i\int_0^{t_1} H(t) dt}|\rm{A}\rangle$
obtained after this operation would still belong to this subspace. Yet,
in a physical system with CH-type interactions, such operations would
necessarily involve relaxation (energy can be exchanged with the
environment and via the magnetic field so that the system relaxes to
its ground state) and decoherence. The relaxation rate(s) in the system
depend(s) on its details, but an important factor on which one can focus
within the CH model is the energy difference between the central branch
and the more conventional single-spin excitations, i.e., spin
waves. Indeed, when one applies an imperfect field with the aim to flip
a whole column, one spin of the column can remain unflipped, or a spin
of a neighboring column can be unintendedly flipped; in both cases this
schematically corresponds to single-spin excitations, as the spin waves
seen in Sec.~\ref{sec:sw}. But if the gap to spin waves is large enough,
these processes should have a negligible impact on the evolution of the
system at moderate time scales. In consequence, we will impose, as a
criterion for the robustness of such a scheme to relaxation
processes, that the lowest single-spin excitation has an energy $E_a$
higher (if possible, much higher) than the energy $E_c^+$ of the highest
eigenstate in the central branch, containing the work subspace.

\begin{figure}[t!]
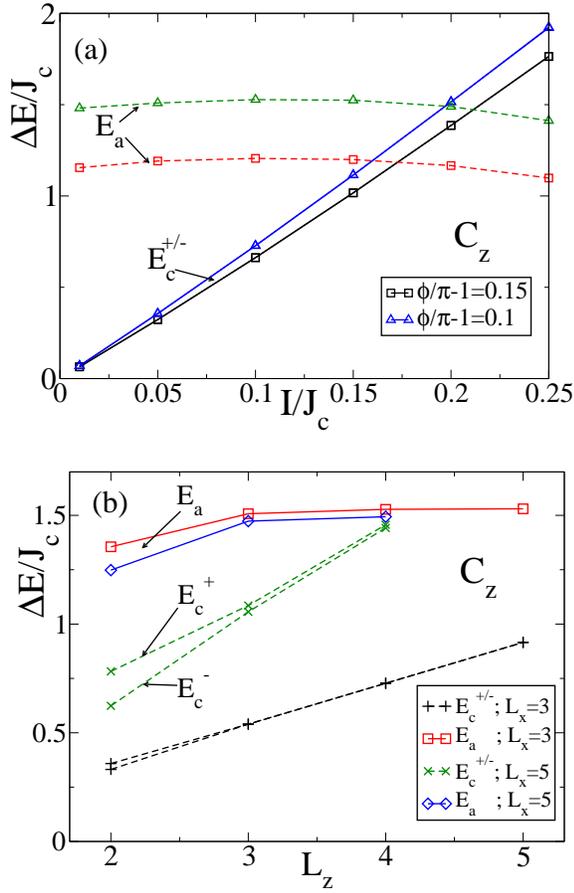

\begin{center}
\includegraphics[width=7.5cm]{CSW_Lx3y4_I.eps}\\
\vspace{0.5cm}
\includegraphics[width=7.3cm]{CSW_I01t23.eps}\\
\caption{\label{fig:swcomp} (Color online) Lowest single spin-flip
excitation energy $E_a$ in the $C_z$-phase in comparison with the
maximal ($E_c^+$) and minimal ($E_c^-$) energies of the \textit{central}
column-flip excitation branch (see text) for open clusters:
(a) as function of $I/J_c$, for fixed $L_x=3$, $L_z=4$, and either
$\phi=11\pi/10$ or $\phi=23\pi/20$ --- here $E_c^+$ and $E_c^-$ are
undistinguishable;
(b) as function of $L_z$, for fixed $I/J_c=0.1$, $\phi=11\pi/10$, and
either $L_x=3$ or $L_x=5$.}
\end{center}
\end{figure}

We examine the dependence of both excitation energies first on
interaction parameters, and second on cluster dimensions $L_x$ and
$L_z$. Figure \ref{fig:swcomp}(a) shows the dependence of $E_c^+$ (and
of $E_c^-$, excitation energy to the lowest eigenstate in the same
branch) on $I/J_c$ for cluster parameters $(L_x=3,L_z=4)$
and two distinct values of $\phi$. In these examples, even for the
largest value of $J_x/J_z$ (i.e., $\phi=23\pi/20$) the dispersion
within this branch is negligible with respect to its lowest excitation
energy ($|E_c^+-E_c^-| \ll E_c^-$); the linear dependence of $E_c^{\pm}$
on $I$ and the weak dependence on $\phi$ are consistent with the fact
that these are columnar excitations, described by the pseudospin
formalism of Sec.~\ref{sec:eff}. In contrast the gap $E_a$ to
single-spin excitations is much more sensitive to $J_x/J_z$; it is
reduced when interactions become less anisotropic, as in the bulk case
discussed in Sec.~\ref{sec:sw} - and has only weak dependance on $I/J_c$
in the regime considered. We stress that $E_a$ is, as in Sec.
\ref{sec:sw}, the lowest single-spin excitation, but here due to open
boundaries it is roughly twice as small as the value expected, with
identical interaction parameters, from Eq.~(\ref{eq:swcz}). Indeed, open
boundaries allow for edge modes of lower energy than the bulk spin wave
modes, since flipping a spin at one (horizontal) edge frustrates only
one $z$-bond compass coupling instead of two.

The dependence of column-flip excitations on cluster dimensions is
displayed in Fig.~\ref{fig:swcomp}(b). The figure shows for fixed
interaction parameters the values of $E_c^+$, $E_c^-$, and $E_a$ as
function of $L_z$ for two values $L_x=3$ and $5$ compatible with the
scheme described in Sec.~\ref{sec:qccol}. The linearity of $E_c^\pm$ in
$L_z$, expected from perturbation theory when assuming
$|I|,|J_x|\ll |J_z|$, is clearly verified. The slope
$dE_c^\pm/dL_z$ is about twice as large for $L_x=5$ as for
$L_x=3$. Indeed, in the first case the central branch is built on
columnar states which contain two domain walls (i.e., in which
$\tau_i^z \tau_{i+1}^z=1$ for exactly two values of $i=1,..,L_x-1$),
while in the latter case the corresponding branch is built on
columnar states with only one domain wall. In contrast, we see that
$E_a$ is roughly size-independent (only for $L_z=2$ its value differs
noticeably from those for longer columns), and reduced of $\simeq 25\%$
from the value $2|J_z|$ expected in the $|I|,|J_x| \ll |J_z|$ limit ---
an adaptation of LSW theory to finite clusters could provide a closer
estimation taking
into account the effect of Heisenberg and $x$-compass couplings.

Based on these features, we define a \textit{column-flip regime} where
the columnar excitations corresponding to the central branch have lower
energy than the lowest single-spin excitation, i.e., $E_c^+<E_a$; the
complementary case is called \textit{spin-wave regime}. In Fig.
\ref{fig:swcomp}(a) we see that the extent of the column-flip regime, in
terms of $I/J_c$, is decreased with increasing $J_x/J_z$ for
$(L_x,L_z)=(3,4)$ --- this regime corresponds to $I/J_c \leq 0.20(1)$
for $\phi=11\pi/10$ and to $I/J_c \leq 0.17(1)$ for $\phi=23\pi/20$.
Increasing cluster dimensions reduces the extent of the column-flip
regime in parameter space, due to the strong dependence, seen in
Fig.~\ref{fig:swcomp}(b), of $E_c^+$ on both $L_x$ and $L_z$ parameters.

Importantly, the column length has also a strong impact on $E_c^+-E_c^-$:
if e.g. $L_x=3$, the four states of this branch are quasi-degenerate
(the splitting is not visible to the eye) for $L_z\ge 3$, while they are
split by energies of $\lesssim 0.05 J_c$ for $L_z=2$. The same trend is
observed for $L_x=5$, but the higher number of states in that branch
results in a significantly broader dispersion, $\simeq 0.2J_c$ for
$L_z=2$ and decreasing rapidly with increasing column length but still
visible in the figure for $L_z=3,4$. This is in agreement with the
perturbation theory approach in Sec.~\ref{sec:eff} according to which
the splitting within a branch, governed by parameters $C_x$ and $C_y$,
decreases exponentially with increasing $L_z$.

We can now make use of these results and formulate the conditions for
such a system to be in the column-flip regime, and simultaneously
require that the dispersion $E_c^+-E_c^- \ll E_c^-$, so that related
decoherence effects are as small as possible. The first condition
requires rather short columns (because of the scaling
$E_c^\pm \propto L_zI$), while the second condition requires
sufficiently long columns [$L_z>2$ with the interaction parameters
chosen for Fig.~\ref{fig:swcomp}(b)]. In the $L_x=3$ case, which might
allow us to engineer a one-qubit device, for values $L_z=3,4,5$ both
conditions can be fulfilled; but in the $L_x=5$ case, which would
correspond to a two-qubit device, these conditions are fulfilled only
for $L_z=3$. Of course, the scaling $E_c^\pm \propto L_zI$ implies that
not only the column length but also the amplitude of Heisenberg
couplings must be rather small; and from Sec.~\ref{sec:sw} we know that
small values of $J_x/J_z$ (i.e., strongly anisotropic compass couplings)
are preferable, to keep these excitation energies below those of
single-spin flips. Thus, in principle, one can engineer clusters with
large dimensions, i.e., large number of qubits, based on
realizations of the
CH model where $J_x/J_z$ and $I/J_z$ are tunable at wish.

We now consider these conditions with a specific physical system in
mind: arrays of quantum spins with dipolar interactions (for
instance, representing NV centers as discussed in Sec.~\ref{sec:int}).
There, the choice of interaction- and size parameters is restricted
further. If we neglect the role of interactions beyond nearest neighbor
effective spins, \cite{lri} the Heisenberg-type couplings on vertical
bonds and on horizontal bonds have distinct (negative) amplitudes, $J_z/3$
and $J_x/3$ respectively, instead of a uniform amplitude $I$; and the
geometry of the array fixes the anisotropy ratio $J_x/J_z$ to
$\zeta^3$, with $\zeta=c/a \in (0;1)$ the ratio between vertical ($c$)
and horizontal ($a$) bond lengths. Note that $\zeta$ is the only
parameter governing all ratios between the various coupling amplitudes.

One can imagine an array of spins with dipolar couplings, restricted by
some screening mechanism to nearest neighbors; and such that, as in
Fig.~\ref{fig:swcomp}(b), $I_x/J_x \simeq -0.33$ (with $I_x$
the Heisenberg-type amplitude on horizontal bonds) and
$J_x/J_z \simeq 0.33$. The corresponding aspect ratio would be
$\zeta=0.69$, but the main difference between this situation and the
case of Fig.~\ref{fig:swcomp}(b) would reside in the amplitude of
Heisenberg-type couplings on vertical bonds ($I_z\simeq 0.3J_c$ in the
dipolar case compared to $I=I_z=I_x\simeq0.1J_c$ in the CH case). This
difference should not play a significant role on values of $E_c^\pm$
since the corresponding states of this branch have ferromagnetically
aligned columns; in fact, in the dipolar case with $\zeta=0.69$, we checked
that the gap to lowest single-spin excitations would then be reduced
compared to $E_a$ seen in Fig.~\ref{fig:swcomp}(b), by a factor of
$\sim 1.5$ only, and branches of quasi-degenerate excitations would
still be present in the low-energy spectrum of e.g. the $(L_x,L_z)=(3,4)$
cluster. Thus, in principle, an array with this or smaller values of
$\zeta$ would allow to define a one- or two-qubit system, assuming that
such highly anisotropic arrays, with dominant dipolar interactions, are
conceivable on an experimental point of view.

\section{Summary and discussion}
\label{sec:summa}

In this work we have investigated the ground states and elementary
excitations of the compass-Heisenberg model formed by quantum spins.
The compass model is characterized by a macroscopic ground state
degeneracy, therefore one of our central goals is to explore what
happens with this huge degeneracy when the system is not perfect but is
exposed to other perturbing interactions, which we assume here to be
of Heisenberg type. Compass interactions can arise in the description
of a variety of strongly correlated electronic systems, ranging from
orbitally degenerate Mott insulators to cold atoms or ions in optical
lattices, the latter having attracted attention in the quest for
possible realizations of quantum computing devices. Besides, we pointed
out that the compass-Heisenberg model can be seen as a short-ranged
version of a Hamiltonian for NV centers, coupled by dipolar interactions,
and also studied intensively recently with motivations from quantum
information.

We first analyzed the zero-temperature phase diagram of this model,
using analytical approaches and numerical exact diagonalization of the
Hamiltonian. We have found that a feature characteristic of the compass
model, the semi-macroscopic ground state degeneracy in the thermodynamic
limit (TL) is lifted in presence of Heisenberg interactions,
even when their amplitude is infinitesimally small.
As a result, the phase diagram contains various ordered phases,
either with ferromagnetic or antiferromagnetic order.
The latter include different
columnar ordered phases. Due to the anisotropy of interactions in spin
space, i.e., except for special values of the parameters $I/J_z$ and
$J_x/J_z$, these phases have an easy spin axis, and the ground state
degeneracy in the TL is twofold. Transitions between
these phases occur for coupling amplitudes either equal or very close
to the corresponding values in the classical (large spin) limit of the
model --- in the first case, this apparent insensitivity to quantum
fluctuations results from extra symmetries of the Hamiltonian at
particular transition lines. In the second case, we presented a rather
precise perturbative evaluation of quantum corrections to the phase
boundaries, consistent with the shifts obtained by exact
diagonalization.

The phase transitions in the compass-Heisenberg model are continuous on
the finite systems studied; tentative size scalings indicate that they
may become of first order in the TL, but alternatively they may keep a
continuous character, since they are characterized by the presence of
gapless excitations in contrast to the ordered phases selected elsewhere.
The modes becoming soft at the transitions can be of two types: for
those occurring at finite $I$ these are spin waves, for which the
dispersion in ordered phases can be well described in the linear
approximation. The case of transitions at $I=0$ between two magnetically
ordered phases is specific: such a transition corresponds to a level
crossing between a multitude of columnar states, which allow us to define
excitations characteristic of the compass-Heisenberg model.

These \textit{column-flip excitations} consist of flips of all spins
within a column, assuming that the dominant compass couplings are those
on vertical bonds. The corresponding excited states, which belong to
the subspace spanned by ground states of the compass model at $I=0$,
are for small but finite $I$ split off proportionally to this amplitude
and to the size of columns. This also implies that they are pushed up to 
high energies in the TL. In small nanoscale systems, however, column flips
can be the lowest excitations, i.e., they may lie inside the anisotropy
gap of spin waves. Column-flip excitations also have the remarkable
property of being grouped into multiplet-branches of quasi-degenerate
states --- the splitting within a branch decreases exponentially with
increasing column length and the number of states in certain branches
grows exponentially with the number of columns. These features, and more
generally the properties of these excitations, are described in detail
by an effective 1D model, which we derived in high-order perturbation
theory.
The effective 1D Hamiltonian couples nearest neighbor $\tau=1/2$ spins,
with terms of the XYZ-type.
We suggest that this situation might be realized in some
transition metal oxides; for instance similar effective 1D
interactions and multiplets with high degeneracy occur also
in the model for manganites.\cite{Lia11}

The effective low-energy Hamiltonian allows, in the regime where this
perturbation theory applies, to describe the quantum dynamics of a
finite cluster of $L_x \times L_z$ spins, with transitions from the
ground state to excited \textit{columnar states} and between the latter
ones. Based on this, we propose a novel type of quantum computing
device, where qubits are physically encoded in specific columns of a
cluster, and where the \textit{work subspace} is embedded in a
quasi-degenerate excitation branch. The encoding in columns instead of
single spins renders the system fault tolerant, similar to proposals
for topological quantum computing --- although a non-trivial topology
is not required here. We suggest here a possible realization of this
type of encoding by means of rectangular arrays of quantum spins
(each spin  representing e.g. a NV center in a diamond matrix)
coupled by dipolar interactions rapidly decaying with distance.
The truncation of these interactions to first neighbors represents, up
to minor details, a particular case of the compass-Heisenberg model
with {\it ferromagnetic} compass couplings. Thus the peculiar features
of low-energy excitations in the latter model could also be encountered
in this more realistic system.

The ferromagnetic nature of dominant compass couplings presents an
advantage for such a realization, as it allows us for an easy
manipulation of columns by an external field. A device would have to
fulfil several conditions on the multiplet-branch containing the
\textit{work subspace}
in order to reduce decoherence:
(i) it should be well separated from single-spin excitations, and
(ii) the energy splitting within this branch should be sufficiently
small.
We analyzed these conditions, finding that they impose restrictions on
the dimensions of the array. To satisfy both conditions, an optimum has
to be found for the length of columns, while the number of columns,
determining the number of hypothetically available qubits, has to be
limited to satisfy the former condition. This number can nevertheless
be increased by varying the geometry (namely the aspect ratio $\zeta$)
of the array, although an unrealistic aspect ratio corresponding to
quasi-decoupled columns may lead to other (decoherence-related) problems.

A possible further development of this study would be to simulate the time
evolution of arrays of spins within the compass-Heisenberg model and to
estimate decoherence and relaxation times. This involves to compute the
reduced density matrix corresponding to the pseudospins defining the
qubits of the system, while other degrees of freedom are traced out. The
time evolution can be studied in the framework of the compass-Heisenberg
Hamiltonian itself, where one can also add perturbing terms accounting
for unavoidable noise effects, and it is also possible to model
elementary operations by including time-dependent fields centered on
specific columns. In the latter context the effective Hamiltonian
represents a great advantage as it allows one to simulate the time-evolution
of qubits much more effectively, i.e., compared to the original
compass-Heisenberg model.

Another direction to follow, more closely connected to arrays of NV centers,
would be to consider, instead of the (short-ranged) CH interactions, the
(power-law-decaying) dipolar interactions and reexamine the conditions
for a similar encoding. Eventually, taking a more theoretical point of
view, the apparently contradicting features seen at the quantum phase
transitions occurring at $I\ne 0$ (mode softening in the spin-wave
spectrum, versus size scalings indicating a conventional first order
scenario in the TL) call for further work using complementary approaches,
especially for transitions not characterized by additional symmetry
on the transition line.
%%%%%%%%%%%%%%%%%%%%%%%%%%%%%%%%%%%%%%%%%%%%%%%%%%%%%%%%%%%%
%%                    Acknowledgments
%%%%%%%%%%%%%%%%%%%%%%%%%%%%%%%%%%%%%%%%%%%%%%%%%%%%%%%%%%%%
\acknowledgments

We thank N. Hasselmann for insightful discussions.
A.M.O acknowledges support by the Polish National Science
Center (NCN) under Project No. N202 069639.

%%%%%%%%%%%%%%%%%%%%%%%%%%%%%%%%%%%%%%%%%%%%%%%%%%%%%%%%%%%%
%%                       References
%%%%%%%%%%%%%%%%%%%%%%%%%%%%%%%%%%%%%%%%%%%%%%%%%%%%%%%%%%%%


\begin{thebibliography}{99}
\bibitem{noteco} Within this approach corrections of finite size
                 systems result from the contributions of nearest
                 neighbor and on-site spin correlations --- only at
                 short distances $d \leq 1$ the correlations differ
                 from the long-distance limit, i.e., entanglement
                 between further neighbor spins is neglected at this
                 order in perturbation theory.
\bibitem{lri} These interactions should of course be included in a more
              realistic description of these systems; we checked that the
              inclusion of second- and third-neighbor couplings from
              $H_{\rm dip}$ does not spoil the branch structure of the
              low-energy spectrum in situations typically considered, for
              instance these shown in Fig.~17.
\bibitem{rows} We keep the notion column-flip excitation also in the opposite
               case $|J_z|<|J_x|$ where actually the rows are flipped.
\end{thebibliography}
\end{document}